\documentclass[]{pasj02} 
\usepackage[switch,mathlines]{lineno} % add line number to manuscript
\usepackage[markup=underlined]{changes}
\setaddedmarkup{\textcolor{red}{#1}}
\setdeletedmarkup{\textcolor{gray}{\sout{#1}}}
\usepackage{subcaption}
\usepackage{needspace}

\jyear{2025}
\Received{}%{yyyy/mm/dd}
\Accepted{}%{yyyy/mm/dd}
\begin{document} 

\title{Diversity of Type Ia supernova optical light curves among different spectroscopic subclasses}

%%% begin:list of authors
% Do NOT capitalize all letters in "textsc".
\author{
 Ryotaro B. \textsc{koshi},\altaffilmark{1} \email{koshi@ioa.s.u-tokyo.ac.jp} 
 Mamoru \textsc{doi},\altaffilmark{2}
 Shigeyuki \textsc{sako},\altaffilmark{1}
 Keiichi \textsc{maeda},\altaffilmark{3}
 Masaomi \textsc{tanaka},\altaffilmark{4,5} and
 Naohiro \textsc{takanashi}\altaffilmark{6}
}
\altaffiltext{1}{Institute of Astronomy, Graduate School of Science, The University of Tokyo, 2-21-1 Osawa, Mitaka, Tokyo 181-0015}
\altaffiltext{2}{National Astronomical Observatory of Japan, 2-21-1 Osawa, Mitaka, Tokyo 181-8588}
\altaffiltext{3}{Department of Astronomy, Kyoto University, Kitashirakawa-Oiwake-cho, Sakyo-ku, Kyoto 606-8502}
\altaffiltext{4}{Astronomical Institute, Graduate School of Science, Tohoku University, Sendai 980-8578}
\altaffiltext{5}{Division for the Establishment of Frontier Sciences, Organization for Advanced Studies, Tohoku University, Sendai 980-8577}
\altaffiltext{6}{Executive Management Program, The University of Tokyo, 7-3-1 Hongo, Bunkyo-ku, Tokyo 113-8654}

%%% end:list of authors

%% !!! Select 3 to 5 words from PASJ's key words !!! 
%% List of Key Words: https://academic.oup.com/pasj/pages/Pasj_Keywords 
%% "\KeyWords{ }" always has to be placed before ``\maketitle'' 
\KeyWords{supernovae: general --- method: data analysis --- surveys}  

\maketitle

\begin{abstract}
Attempts to reveal the spectroscopic diversity of Type Ia supernovae (SNe Ia) have led to subclassification schemes such as the Branch system, which classifies SNe Ia into four categories: core normal (CN), broad line (BL), cool (CL), and shallow silicon (SS). The physical origin of these spectroscopic differences, including progenitor channels, explosion mechanisms, or other parameters, however, remains unclear. Moreover, previous work has concentrated primarily on properties near peak luminosity, yielding limited insight into their behavior at later epochs. In this study, we compile $UBVRI$ photometry for 109 SNe Ia and construct the first set of average light curves for each Branch subgroup, spanning from pre-maximum through the late tail. We find pronounced diversity in the $I$-band, especially in the timing of the secondary maximum across subgroups and in the late-time decline of CL events. After correcting for light curve stretch, which reflects the combined influence of ${}^{56}$Ni and ejecta masses, we show that the secondary maximum is powered by Fe II recombination, and its timing is particularly sensitive to the amount of stable iron-group elements (IGEs) synthesized in the explosion. This implies an anti-correlation between the mass of stable IGEs and ${}^{56}$Ni: CL (and possibly BL) events have a larger mass ratio of IGEs/$^{56}$Ni resulting in earlier secondary maxima, while SS events have a smaller ratio and thus later secondary maxima. This trend is naturally explained within the near-$M_{\rm Ch}$ delayed-detonation scenario, whereas it is inconsistent with the positive correlation predicted by the sub-$M_{\rm Ch}$ double-detonation scenario. Finally, we show that stretch-corrected late-time slopes provide a practical diagnostic for CL events, likely linked to an emission feature around $7,200$ \AA.
\end{abstract}

%\pagewiselinenumbers 

\section{Introduction}\label{sec:introduction}
Type Ia supernovae (SNe Ia) are thermonuclear explosions of carbon-oxygen white dwarfs (WDs) that exhibit a characteristic Si II absorption feature in their optical spectra at maximum light. SNe Ia are widely believed to arise from a mass-accreting white dwarf in a binary system, but the nature of the companion star remains uncertain. Two leading scenarios are considered today; in the single-degenerate scenario (e.g., \cite{whelan73}; \cite{nomoto82}), a non-degenerate star transfers mass onto the WD until it ignites, whereas in the double-degenerate channel (e.g., \cite{iben84}; \cite{webbink84}), the merger of two WDs triggers the explosion. 

Besides the progenitor problem, the explosion mechanism itself remains under debate and is an independent factor in understanding the nature of SNe Ia. In classical models, deflagration starts near the center of the WD when the density reaches $\sim{10}^9$ ${\rm g}$~${\rm cm}^{-3}$ and then shifts to detonation (delayed-detonation model; \cite{woosley95}; \cite{iwamoto99}; \cite{seitenzahl13}). However, an alternative model claims that detonation can occur at lower WD masses and densities, triggered by an initial detonation at the surface helium-rich layer (double-detonation model; \cite{kromer10}; \cite{shen14}; \cite{polin19}). 

Recent multi-dimensional simulations have further highlighted that, in the delayed-detonation model, the relative strengths of deflagration and detonation are antagonistic: stronger initial deflagration leads to greater pre-expansion, which reduces the efficiency of the subsequent detonation and lowers the $^{56}$Ni yield (\cite{krueger10}; \cite{maeda10c}; \cite{lach22}). In addition, enhanced burning at high densities during strong deflagration promotes electron captures, resulting in larger amounts of stable iron-group elements such as $^{58}$Ni and $^{54}$Fe (\cite{blondin22}).

Note that the explosion mechanism is not uniquely determined by the progenitor channel; even with the companion identified, how ignition and detonation proceed remains uncertain (see \cite{maeda16} for a review). Connecting observations to theoretical models provides crucial constraints on both fronts, making the observed diversity of SNe Ia a central theme in current research.

Studies of photometric diversity have long underpinned our understanding and standardization of SNe Ia. The stretch–magnitude relation is an empirical correlation between the stretch of the light curve (or in some studies, the decline rate) and peak absolute magnitude, showing that slowly declining SNe Ia are intrinsically brighter (\cite{phillips93}). Another correlation stands between the color at peak luminosity and the peak absolute magnitude; bluer SNe Ia appear more luminous (\cite{tripp98}). These relations both potentially illuminate underlying explosion physics, such as the mass of synthesized ${}^{56}$Ni, and enable calibration of SN Ia luminosity, establishing their use in observational cosmology (\cite{riess98}; \cite{perlmutter99}). More recent cosmological analyses, such as the Pantheon and Pantheon+ samples, have confirmed and refined these relations using thousands of SNe Ia (\cite{scolnic18}; \cite{brout22}).

In addition to these photometric diagnostics, spectroscopic diversity also provides crucial insight. \citet{branch06} defined four subgroups of SNe Ia based on the pseudo-equivalent widths (pEWs) of the Si II $\lambda6355$ and Si II $\lambda5972$ absorption lines near maximum light: core normal (CN) events, which exhibit intermediate line strengths and represent the archetypal normal population; broad line (BL) SNe Ia, distinguished by strong Si II $\lambda6355$ absorption and broader profiles indicative of higher ejecta velocities; cool (CL) SNe Ia, characterized by a high $\lambda5972/\lambda6355$ ratio and generally subluminous behavior (including the 1991bg-like class); and shallow silicon (SS) SNe Ia, in which both Si II lines are weak, corresponding predominantly to the overluminous 1991T-like/1999aa-like subclass (see figure 1 of \cite{branch09} for representative spectra of each subtype). \citet{wang13} divided spectroscopically normal SNe Ia into two subgroups based on the Si II $\lambda6355$ velocity near maximum light: the normal velocity (NV; $v\lesssim11,800{\rm km}{\rm s}^{-1}$) and high velocity (HV; $v\gtrsim11,800{\rm km}{\rm s}^{-1}$) groups, which exhibit differences in color and host galaxy properties (see also \cite{benetti05}). The Wang NV and HV groups vaguely overlap with the Branch CN and BL groups respectively. 

These subclassification schemes are not merely empirical. CN events are broadly consistent with the standard Chandrasekhar-mass delayed-detonation model (\cite{seitenzahl13}; \cite{blondin13}), while BL/HV events are often interpreted as having more extended outer ejecta structures or stronger asymmetry, possibly linked to off-center ignition (\cite{mazzali05}; \cite{stehle05}; \cite{maeda10a}) or the double-detonation model (\cite{ogawa23}). CL events point to reduced $^{56}$Ni production and cooler ejecta (\cite{nugent95}; \cite{taubenberger08}; \cite{dhawan17}). SS events, on the other hand, are associated with more extended burning or strong mixing (\cite{ruiz-lapuente92}; \cite{sasdelli14}). Although no subtype can yet be tied to a unique progenitor channel or explosion mechanism, these patterns suggest that spectroscopic classifications carry physical significance, providing essential clues to the diversity of SNe Ia.

Efforts to link photometric and spectroscopic properties have focused primarily on peak‐epoch observables. \citet{benetti05} showed that $\Delta m_{15}({\rm B})$, defined as the difference in $B$‑band magnitude between maximum light and 15 days afterward, correlates with the pEW ratio of the two Si II lines, and \citet{hachinger06} extended this to additional line ratios and the pEW of $\lambda$5972, all of which correlate with peak luminosity. \citet{folatelli13}, using 93 SNe Ia from the Carnegie Supernova Project, further quantified correlations between $\Delta m_{15}(B)$, pEWs, pEW ratios, and line velocities at maximum light. Colors of SNe Ia at peak are also found to have correlations with the pEW of Si II $\lambda6355$ \AA, with lower velocities corresponding to bluer colors (\cite{foley11}). More recent studies have begun to explore multi-epoch spectra; for instance, the time evolution of Si II, Ca II, O I, and S II features in early to near-peak spectra have been systematically analyzed (\cite{zhao15}; \cite{zhao16}; \cite{zhao21}). However, comparisons that incorporate information from a wider range of epochs remain scarce, even though they promise a more comprehensive view of SN composition and explosion history. Existing investigations have largely emphasized the evolution of specific line diagnostics within limited phase ranges, leaving open the broader connection between spectroscopic diversity and photometric behavior.

One compelling example is the secondary maximum seen in the $I$‐band (and other NIR bands) 20–30 days after $B$‐band peak. \citet{kasen06} demonstrated that, as the ejecta cools to $\sim$7000 K, iron‐group elements (IGEs) recombine and an inward‐receding recombination front enhances I‐band emissivity, producing the secondary peak. Variations in its timing and strength should reflect differences in physical properties such as ${}^{56}$Ni mass, mixing, and the synthesis of stable iron. On the observational side, \citet{jack15} identified a blended Fe II spectral feature around 7,500 \AA~in SN 2014J coinciding with the onset of the secondary maximum, providing empirical support for the theoretical mechanism. Observations have revealed diversity of this phase as well. \citet{takanashi08}, hereafter TAK08, made $UBVRI$ average light curves from 122 nearby SNe Ia and found that diversity was seen in the I-band secondary maximum, and that most of the outliers had peculiar spectroscopic properties. More recently, \citet{deckers25} analyzed nearly 900 SNe Ia from the Zwicky Transient Facility (ZTF) DR2 sample and statistically found correlations between the timing and strength of the $r$-band secondary maximum and $\Delta m_{15}(g)$. In the $i$-band, the timing showed a similar correlation, whereas the strength exhibited a larger scatter. While their study provided valuable empirical relations based purely on photometry, the underlying physical mechanism remains unclear. Some studies have found that some CL SNe Ia lack a distinct secondary maximum, which is thought to be due to low temperatures (\cite{turatto96}; \cite{dhawan17}). The secondary maxima of HV SNe Ia have also been found to show different behaviors from NV SNe Ia. \citet{burgaz21} found that HV SNe Ia have a larger scatter in magnitude and being brighter on average, and linked the luminosity difference to the distribution of $^{56}$Ni in the ejecta; that HV may have an ejecta structure with more $^{56}$Ni spread out in the outer regions. These results emphasize the need to link photometry and spectroscopy at various phases in their evolution.

In this work, we construct average $UBVRI$ light curves for each Branch subtype to investigate how photometric evolution differs among spectroscopic classes. We then combine these light curve templates with spectral diagnostics to explore the underlying physical drivers of SN Ia diversity beyond maximum light. Section 2 describes our data sample and processing; section 3 details the creation of mean light curves; section 4 presents fitting results and the average light curves; and section 5 discusses the implications for progenitor systems and explosion physics.

\section{Data}\label{sec:data}
\subsection{Archival data}
\begin{figure}[htbp]
 \begin{center}
  \includegraphics[width=8cm]{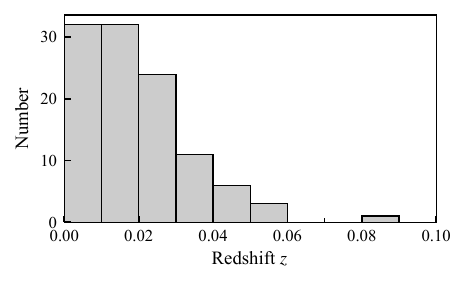} 
 \end{center}
\caption{Histogram of the redshift distribution of the SN Ia sample used in this work. Bins have a width of $\Delta z=0.01$. {Alt text: The x axis shows the redshift from 0.00 to 0.10. The y axis shows the number of supernovae, from 0 to 35.}}\label{fig:z_dist}
\end{figure}

The optical light curves of SNe Ia used in this study were obtained from two archival sources: the dataset compiled by TAK08, and data from the Carnegie Supernova Project (CSP; \cite{krisciunas17}). TAK08 assembled their sample as Johnson-Cousins $UBVRI$-band (\cite{bessell90}) light curves from various papers (primarily \cite{jha06}). In addition to the original TAK08 compilation, we incorporated several SNe~Ia from the literature (SN~1990N, SN~1997cn, SN~1998aq, SN~1998bp, SN~1999by, SN~2002er, and SN~2003du). Because these objects were also observed in the same $UBVRI$ photometric system, we treat them as part of the TAK08 sample throughout this work. CSP data were acquired as $uBVri$ system photometry (\cite{stritzinger11}) using Swope 1-m telescope, and the Swope 2.5-m du Pont telescope at the Las Campanas Observatory (\cite{krisciunas17}). From these datasets, we selected SNe Ia that had been classified into Branch subgroups in the literature (\cite{branch09}; \cite{folatelli13}), and had at least one data point in all five bands. All objects in our sample are at low redshift (figure \ref{fig:z_dist}). This yields a total sample of 109 SNe Ia, with 37 from the TAK08 sample and 72 from CSP. The number of SNe Ia used in this study is limited by the availability of public datasets that provide both multi-band light curves and high-quality spectra across multiple epochs. In practice, the combination of the TAK08 compilation and the Carnegie Supernova Project constitutes the principal open resources that meet these requirements, and our sample is therefore drawn from these surveys.

We also examine optical spectroscopic properties to further explore the diversity among the Branch subgroups. The spectra employed in this analysis are taken from \citet{blondin12} and \citet{folatelli13}. The values of pEWs are taken from \cite{branch09} for the TAK08 sample, and \cite{folatelli13} for the CSP sample.

\begin{figure}[htbp]
 \begin{center}
  \includegraphics[width=8cm]{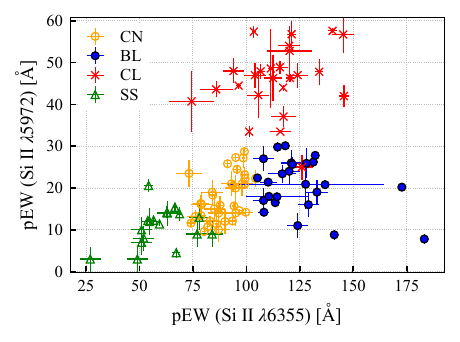} 
 \end{center}
\caption{Branch diagram showing pseudo-equivalent widths of Si II $\lambda 6355$ versus Si II $\lambda 5972$ for the SN Ia sample used in this work. Different subgroups are denoted by symbols: CN (orange open circles), BL (blue filled circles), CL (red crosses), and SS (open triangles). Values are from \citet{branch09} and \citet{folatelli13}. Objects that lack measurements of either Si II line are not plotted here, but are retained for analyses that do not require pEWs. {Alt text: A scatter plot diagram. The x axis shows the pseudo-equivalent width of the Si II absorption line at 6355 Angstroms, and the y axis shows those of the Si II line at 5972 Angstroms.}}\label{fig:branch_diagram}
\end{figure}

Figure \ref{fig:branch_diagram} displays the distribution of the pEWs of the two Si II absorption features for our sample. In this diagram, the different Branch subgroups are generally well separated. One notable exception is SN 1989B, which is classified as CL yet falls in the region predominantly occupied by BL SNe. Despite its weak Si II $\lambda5972$ feature, its overall spectral resemblance to CL justifies its assignment to that subgroup (see \cite{branch09}). Table \ref{tab:n_branch} summarizes the number of SNe in each Branch subgroup.

\begin{table}[htbp]
  \tbl{The number of SNe Ia in each Branch subgroup.}{%
  \begin{tabular}{cccc}
      \hline
      Subgroup & Number & Poor fit\footnotemark[$*$] \\ 
      \hline
      Core normal (CN) & 35 & 1 \\
      Broad line (BL) & 28 & 2 \\
      Cool (CL) & 26 & 13 \\
      Shallow silicon (SS) & 20 & 0\\
      \hline
    \end{tabular}}\label{tab:n_branch}
\begin{tabnote}
\footnotemark[$*$] The number of SNe Ia that yielded poor fits to the initial template (generic average light curves from TAK08), mostly due to peculiar $I$-band light curve shapes. Discussed in detail in section \ref{ssec:poor_fit}.
\end{tabnote}
\end{table}

\subsection{Light curve correction}
As mentioned earlier, the light curves from TAK08 and CSP use different filter sets. The filter sets differ in wavelength range and transmission curves. Moreover, because SNe Ia at different redshifts have their observed photometry shifted to different wavelengths, it is necessary to apply a band correction to each data point prior to the construction of average light curves.

Our correction is based on the cross-filter K-correction method described by \citet{kim96}. In brief, we compute synthetic magnitudes for a template spectrum in both the observed and rest-frame filter systems using the following equation:
\begin{eqnarray}
    K_{\rm XY}&=&-2.5\log_{10}\left[\frac{1}{1+z}\cdot\frac{\int\lambda F_\lambda\left(\frac{\lambda}{1+z}\right)Y(\lambda){\rm d}\lambda}{\int\lambda F_\lambda(\lambda)X(\lambda){\rm d}\lambda}\right. \nonumber\\
    &&\qquad\qquad\qquad\left.\times\frac{\int\lambda S_\lambda^{\rm X}(\lambda)X(\lambda){\rm d}\lambda}{\int\lambda S_\lambda^{\rm Y}(\lambda)Y(\lambda){\rm d}\lambda}\right],
    \label{eq:band_corr}
\end{eqnarray}
where $K_{XY}$ is the correction required to convert from the observed band $Y$ to the rest-frame band $X$, $z$ is the redshift, $F_\lambda(\lambda)$ is the flux density of the template spectrum, $X(\lambda)$ and $Y(\lambda)$ are the filter transmission functions, and $S_\lambda^{\rm X}(\lambda)$ and $S_\lambda^{\rm Y}(\lambda)$ represent the spectral energy distributions of the reference star used to define the zeropoints in each band.

The choice of spectral templates is motivated by their representative nature and widespread use. The \citet{hsiao07} spectra provide an empirical mean sequence of normal SNe Ia and are widely used for K-corrections and light curve modeling, making them appropriate for the CN and BL groups. Although high velocity features make BL spectra distinct from CN spectra, especially in the rising phase, dedicated BL templates are not currently available, and the Hsiao template provides a reasonable proxy, particularly as our analysis focuses on post-maximum phases where CN and BL spectra are broadly similar. For the CL and SS groups, the SN 1991bg-like and SN 1991T-like templates based on \citet{nugent02}\footnote{https://c3.lbl.gov/nugent/nugent\_templates.html} are standard references that capture the distinctive spectral characteristics of these subclasses. Thus, the adopted set of templates reflects the most representative and publicly available resources for each Branch subgroup, ensuring consistency with previous work. Since the light curves from TAK08 are reported in the Vega magnitude system, we used a spectrum of Vega from the European Southern Observatory\footnote{https://www.eso.org/sci/observing/tools/standards/spectra/hr7001.html} to perform the conversion.

Because neither absolute magnitudes nor colors are used in this study, no correction for dust extinction was applied to avoid unnecessary uncertainty in the photometric values. Instead, we use $B-V$ colors as values relative to the color at peak:
\begin{equation}
    \label{eq:color_def}
    (B-V)_{\rm rel}(t)=(B-V)(t)-(B-V)_{\rm peak}.
\end{equation}

For each light curve, the observed magnitude in band $X$, $m_{\rm X}$, is transformed to the rest-frame apparent magnitude in band $Y$, $m_{\rm Y}$, via
\begin{equation}
    \label{mcorr}
    m_{\rm Y} = m_{\rm X} + K_{\rm XY}.
\end{equation}

\section{Method}
We construct average light curves for each Branch subgroup in two main steps: (1) fitting individual SN Ia light curves to a set of templates using a "multi-band stretch method", and (2) averaging the fitted light curves in each subgroup. This approach enables us to separate the effects of stretch, which is a well-studied parameter, from other potential light curve shape properties, and study the intrinsic diversity of SN Ia light curves comprehensively. Each of the following subsections details a part of the process.

\subsection{Light curve fitting}
In our first step, we fit the light curve data to a set of template light curves with a multi-band stretch method described in TAK08. In this method, each photometric band is allowed its own stretch parameter, thereby providing a better overall fit and permitting the investigation of band-specific properties. We adopted the $UBVRI$ average light curves from TAK08 as our initial template. This is a generic light curve made from SNe Ia of different subtypes altogether.

For each SN Ia, we determine a total of 11 parameters: the time of B-band maximum light, denoted as $t_{B, {\rm peak}}$; the stretch factors for each photometric band, represented as $s_Y$; and the peak magnitudes for each band, indicated by $m_Y$, for $Y = \{U, B, V, R, I\}$.

The observed light curve data, consisting of points $(t_{Y,i}, m_{Y,i} \pm \sigma_{Y,i})$, are transformed into the template frame as follows:
\begin{equation}\label{eq:corrections}
\left\{
\begin{array}{l}
    {t'}_{Y,i} = \dfrac{t_{Y,i} - t_{B, {\rm peak}}}{s_Y(1+z)}, \\[6pt]
    {m'}_{Y,i} = m_{Y,i} - m_{Y,{\rm peak}} + m_{\rm temp,peak}.
\end{array}
\right.
\end{equation}
Here, $z$ denotes the redshift, and $m_{\rm temp,peak}$ is the peak magnitude of the template light curve.

For each band, we quantify the difference between the transformed data and the template by computing the weighted chi-square:
\begin{equation}
    \label{chiy}
    \chi^2_Y = \sum_i \frac{\left[ m_{{\rm temp}, Y}\left({t'}_{Y,i}\right) - {m'}_{Y,i} \right]^2}{\sigma^2_{Y,i} + \sigma^2_{{\rm temp}, Y}\left({t'}_{Y,i}\right)}.
\end{equation}
For a fixed value of $t_{B,{\rm peak}}$, we determine the candidate stretch and peak magnitude in each band, $\left[s^{\rm cand}_Y(t_{B,{\rm peak}}),\, m^{\rm cand}_{Y,{\rm peak}}(t_{B,{\rm peak}})\right]$, that minimize $\chi^2_Y$. The overall chi-square is then obtained by summing over all bands:
\begin{equation}
    \label{chi}
    \chi^2 = \sum_Y \chi^2_Y\Big|_{s_Y=s^{\rm cand}_Y(t_{B,{\rm peak}}),\, m_{Y,{\rm peak}}=m^{\rm cand}_{Y,{\rm peak}}(t_{B,{\rm peak}})}.
\end{equation}

The optimal set of parameters, denoted by $\left(t^{\rm fit}_{B,{\rm peak}},\, s^{\rm fit}_Y,\, m^{\rm fit}_{Y,{\rm peak}}\right)$, is obtained by minimizing $\chi^2$. We denote the minimum value as ${\chi^2}_{\rm min}$ and define the reduced chi-square as $\chi^2_{\rm red} = {\chi^2}_{\rm min} / \mbox{dof}$, where “dof” represents the degrees of freedom.

Parameter uncertainties are estimated in the fitting to the initial template using a Markov Chain Monte Carlo (MCMC) approach implemented with the Python package \texttt{emcee} (\cite{foreman-mackey13}). An ensemble of 32 walkers was initialized near the optimal parameter estimates by adding small Gaussian perturbations. The sampler was run for 2000 iterations, with the first 500 iterations discarded as burn-in. The remaining samples were flattened across walkers, and the uncertainty for each parameter was computed as half the difference between the 84th and 16th percentiles of the posterior distribution.

Once the best-fit parameters are determined, each SN Ia light curve is transformed into the template frame (hereafter referred to as a "fit light curve").

\subsection{Making average light curves}
For each Branch subgroup, we compile the fitted light curves to construct an average light curve. Initially, all fitted light curves for a given subtype are overlaid in a time versus magnitude frame to create a master light curve. Next, the data are binned along the time axis, and the weighted average magnitude in each bin is computed using the inverse square of the observational uncertainties as weights.

Because the temporal sampling density decreases substantially at later epochs, we adjust the bin widths as a function of time. For epochs earlier than 40 days after B-band maximum, we adopt a fixed bin width of 3 days. For later epochs, the bin widths are determined by a logarithmic function to ensure that the averaged magnitudes are not overly influenced by a few individual SNe. In addition, consecutive bins are allowed to overlap by half the width of the previous bin to promote smooth transitions in the average light curve.

As a result, we obtain stretch-free average $UBVRI$ light curves for each subgroup. Since all light curves were initially fitted with a common template, the resulting averages still retain a strong dependence on that template. To reduce this bias and better represent the intrinsic properties of each Branch subgroup, we repeat the entire procedure iteratively. In each iteration, the average light curve from the previous cycle serves as the new fitting template. The process continues until the mean squared error (MSE) between the average light curves from successive iterations, ${\rm MSE}_i$, falls below a threshold defined by the average uncertainty of the previous iteration, $\bar{\sigma}_i$:
\begin{equation}
    \label{eq:loop_crit}
    {\rm MSE}_i \leq \bar{\sigma}_i.
\end{equation}

\section{Results}
\subsection{Fitting results}\label{ssec:fit_result}
The best fit parameters and uncertainties for all SNe Ia fit to the initial template are provided in appendix~\ref{app:fit_result} (table~\ref{tab:fit_result}). Hereafter, stretch is defined so that $s=1$ for all bands in the initial template. In the main text, we focus on distribution-level trends and representative correlations.

Throughout this study, we distinguish between two definitions of phase. "Observed phase" refers to the epoch relative to $B$-band maximum without correction, while "stretch-corrected phase" refers to the time axis after dividing it by the fitted stretch factor. Unless otherwise noted, results comparing multiple subtypes (e.g., figures~\ref{fig:average_lc}, \ref{fig:feii_lc_subtype}) are presented in the stretch-corrected phase in order to remove the first-order effect of $^{56}$Ni mass.

\begin{figure*}[htbp]
  \centering
  \begin{minipage}{0.32\textwidth}
    \centering
    \includegraphics[width=\linewidth]{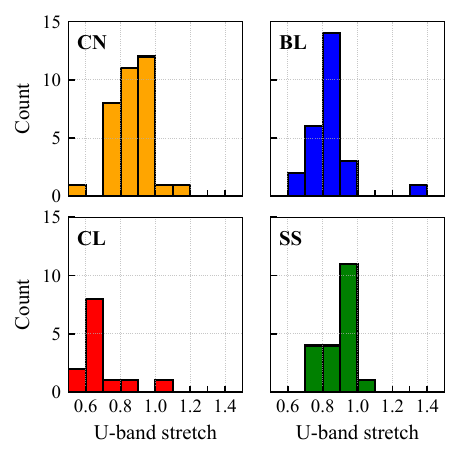}\\
    (a) $U$-band
  \end{minipage}
  \hfill
  \begin{minipage}{0.32\textwidth}
    \centering
    \includegraphics[width=\linewidth]{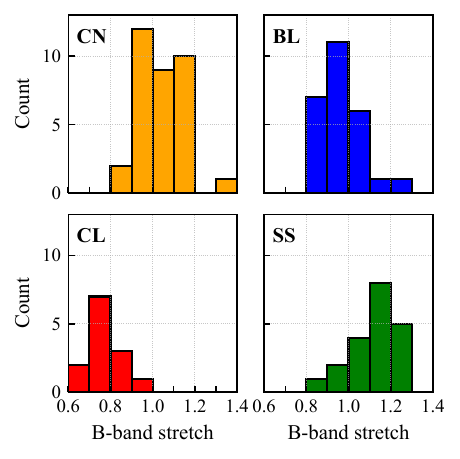}\\
    (b) $B$-band
  \end{minipage}
  \hfill
  \begin{minipage}{0.32\textwidth}
    \centering
    \includegraphics[width=\linewidth]{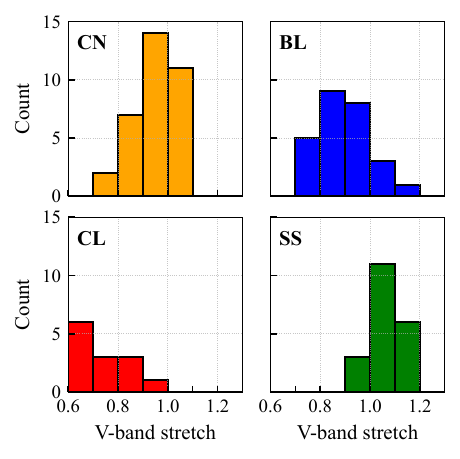}\\
    (c) $V$-band
  \end{minipage}
  \vspace{1.2em}
  \begin{minipage}{0.32\textwidth}
    \centering
    \includegraphics[width=\linewidth]{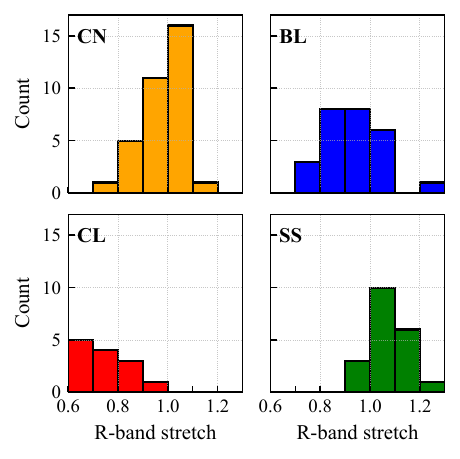}\\
    (d) $R$-band
  \end{minipage}
  \begin{minipage}{0.32\textwidth}
    \centering
    \includegraphics[width=\linewidth]{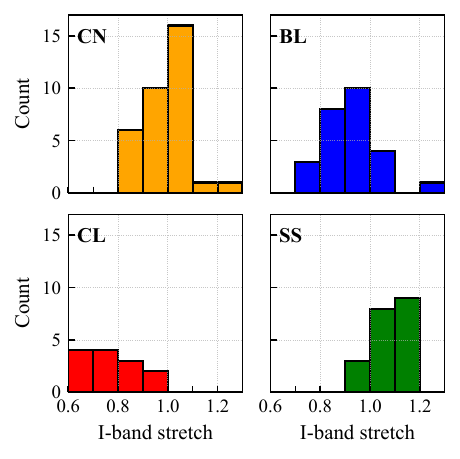}\\
    (e) $I$-band
  \end{minipage}
  \vspace{0.6em}
  \caption{Histograms of stretch distributions for each Branch subtype. Stretch values are defined so that $s=1$ for the initial template. {Alt text: An ensemble of histograms. There are five parts of the figure, corresponding to UBVRI bands. Each part is consisted of four panels, corresponding to the Branch subgroups. Each panel has an x axis showing the stretch value in that band, and the y axis showing the number of supernovae.}}
  \label{fig:stretch_dist}
\end{figure*}

We first inspect the distribution of the derived stretch parameters. Figure~\ref{fig:stretch_dist} shows the distribution of stretch values for well-fit SNe Ia (${\chi_{\rm red}}^2\le4$) in each subtype. Although the subtypes share overlapping ranges of stretch values, their averages differ (table~\ref{tab:stretch_stats}). In all bands, average stretch values increase in the order of CL, BL, CN, and SS.

\begin{table*}[htbp]
  \tbl{Weighted mean $\pm$ 1$\sigma$ of stretch parameters for each Branch subtype.}{%
  \begin{tabular}{lccccc}
      \hline
      \noalign{\vskip 2pt}
      Subgroup & $\overline{s_U^{\rm fit}}$ & $\overline{s_B^{\rm fit}}$ & $\overline{s_V^{\rm fit}}$ & $\overline{s_R^{\rm fit}}$ & $\overline{s_I^{\rm fit}}$ \\ 
      \noalign{\vskip 2pt}
      \hline
      CN & 0.847 $\pm$ 0.084 & 1.012 $\pm$ 0.091 & 0.928 $\pm$ 0.069 & 0.958 $\pm$ 0.097 & 0.949 $\pm$ 0.093 \\
      BL & 0.834 $\pm$ 0.074 & 0.951 $\pm$ 0.076 & 0.863 $\pm$ 0.088 & 0.905 $\pm$ 0.099 & 0.885 $\pm$ 0.088 \\
      CL & 0.660 $\pm$ 0.051 & 0.802 $\pm$ 0.073 & 0.740 $\pm$ 0.085 & 0.740 $\pm$ 0.078 & 0.771 $\pm$ 0.079 \\
      SS & 0.847 $\pm$ 0.095 & 1.146 $\pm$ 0.063 & 1.059 $\pm$ 0.046 & 1.078 $\pm$ 0.054 & 1.091 $\pm$ 0.053 \\
      \hline
    \end{tabular}}\label{tab:stretch_stats}
\begin{tabnote}
\end{tabnote}
\end{table*}

It is well established that CL SNe Ia tend to have small stretch values, while SS SNe Ia exhibit large ones (e.g., \cite{folatelli13}). The distinction between CN and BL, however, has been less clear. \citet{blondin12} reported that BL events have a slightly steeper $B$-band decline rate ($\Delta m_{15}(B)$) than CN, whereas \citet{folatelli13} found no significant difference. Our results provide new evidence by demonstrating that BL consistently shows smaller stretch than CN across all optical bands, not just in the $B$-band.

The stretch parameter represents the characteristic timescale of the light curve, which depends on the opacity, ejecta mass, and ejecta velocity (\cite{arnett82}). The amount of synthesized $^{56}$Ni affects the opacity and thus the diffusion timescale. Observationally, however, the ejecta velocity shows little or no correlation with stretch (\cite{pan24}). This suggests that variations in stretch primarily reflect differences in the $^{56}$Ni and ejecta masses.

\begin{figure}[htbp]
    \centering
    \includegraphics[width=6.0cm]{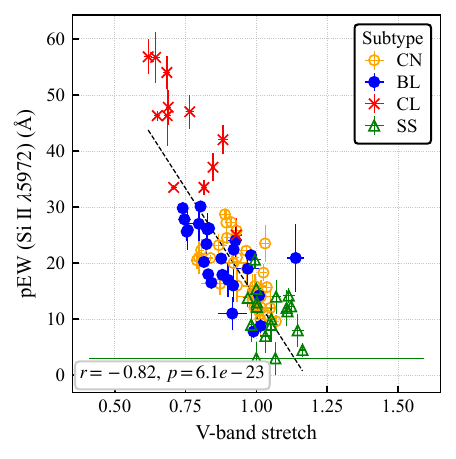}\\[-6pt]
    {\small (a) $V$-band stretch vs. pEW (Si II $\lambda5972$)}\\[4pt]
    \includegraphics[width=6.0cm]{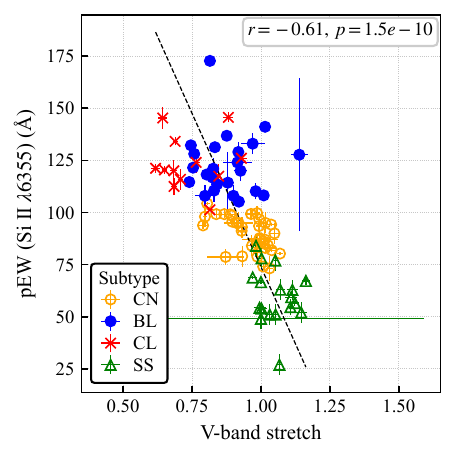}\\[-6pt]
    {\small (b) $V$-band stretch vs. pEW (Si II $\lambda6355$)}\\[4pt]
    \includegraphics[width=6.0cm]{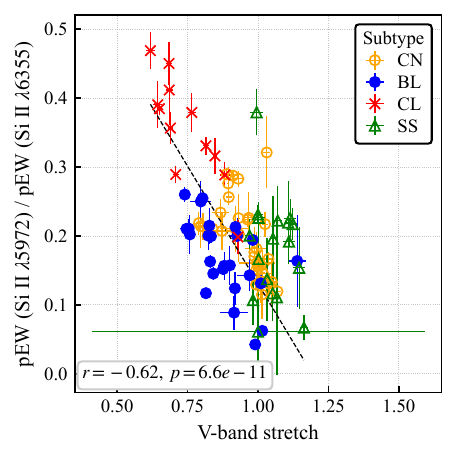}\\[-6pt]
    {\small (c) $V$-band stretch vs. pEW ratio of Si II $\lambda5972$ and $\lambda6355$}\\[3.5pt]
    \caption{Correlation between $V$-band stretch and the pEW of (a) Si II $\lambda5972$, (b) Si II $\lambda6355$, and (c) their ratio. Symbols follow the same convention as in figure~\ref{fig:branch_diagram}. The dashed line in each panel shows the best fit linear relation obtained with orthogonal distance regression. The Pearson coefficient $r$ and $p$-value are also shown. {Alt text: A figure consisting of three panels. All panels have V-band stretch as the x axis. The y axes show the pseudo-equivalent widths of the Si II line at 6355 Angstroms, the Si II line at 5972 Angstroms, and the ratio of the two lines, for each panel.}}
    \label{fig:sv_pew}
\end{figure}

We then examined how the fitted stretch parameters in each band correlate with the pEWs of the Si II $\lambda$5750 and $\lambda$6355 absorption lines and their ratio. Three representative cases that compare $V$-band stretch to the pEWs are shown in figure~\ref{fig:sv_pew}. Across the full sample, light curves with larger stretch tend to have weaker Si~II absorption. The pEW ratio of the two Si II lines also correlate with stretch. Similar results for the other bands and subtype-resolved correlation coefficients are presented in appendix~\ref{app:stretch_vs_pew} 
(figures~\ref{fig:su_pew}-–\ref{fig:si_pew}; table~\ref{tab:corr_pew}). These results are consistent with previous work (e.g., \cite{benetti05}; \cite{folatelli13}), but this work further extends these correlations to other optical bands.

When divided by Branch subtype, however, these inverse stretch–pEW trends are not uniform. CN, BL, and CL retain robust anti-correlations with the Si II $\lambda5972$ feature in every band, whereas SS shows little to no relation, likely at least partially reflecting the difficulty of measuring weak Si~II in SS. For the stronger Si II $\lambda6355$ line, only CN preserves a significant negative slope in B, V, R, and I; BL weakens, and CL/SS are largely uncorrelated. Taking the pEW ratio, pEW (Si II $\lambda5972$) / pEW (Si II $\lambda6355$), sharpens the stretch dependence for CL (reaching $|r|\sim0.8$), while SS remains uncorrelated. These systematic differences by Branch subclass show that stretch values correlate with different spectroscopic parameters across the subclasses. One alternative explanation is that, because each subclass spans only a limited range of stretch values, the correlations appear less pronounced when the groups are analyzed separately. Nevertheless, the difference between CN and BL remains noteworthy, since the widths of their stretch distributions are comparable.

\subsection{Average light curves}
Following the method described earlier, the template light curves for each Branch subtype were constructed by iteratively refining the fit until the residual difference between consecutive iterations converged below a critical threshold. Table~\ref{tab:n_iteration} summarizes the number of iterations performed for each subtype and photometric band.

\begin{table}[htbp]
  \tbl{The number of samples (iterations) used to make the average light curve for each subtype and band.}{%
  \begin{tabular}{lccccc}
      \hline
      Subgroup & $U$-band & $B$-band & $V$-band & $R$-band & $I$-band \\ 
      \hline
      CN & 33 (2) & 33 (2) & 33 (2) & 33 (2) & 33 (2) \\
      BL & 23 (2) & 23 (2) & 23 (2) & 23 (2) & 23 (2) \\
      CL & 15 (7) & 13 (2) & 13 (2) & 11 (9) & 16 (3) \\
      SS & 17 (2) & 17 (2) & 17 (2) & 16 (4) & 17 (2) \\
      \hline
    \end{tabular}}\label{tab:n_iteration}
\begin{tabnote}
\end{tabnote}
\end{table}

\begin{figure*}[htbp]
  \centering
  \begin{minipage}{0.45\textwidth}
    \centering
    \includegraphics[width=8.0cm]{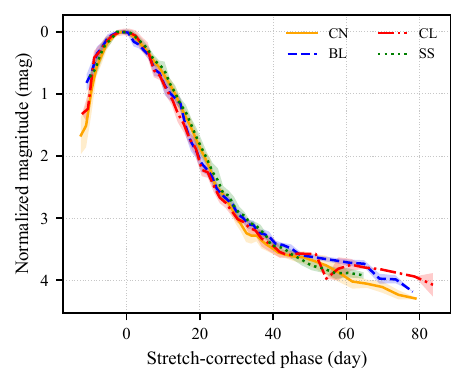}\\
    (a) $U$-band
  \end{minipage}
  \hfill
  \begin{minipage}{0.45\textwidth}
    \centering
    \includegraphics[width=8.0cm]{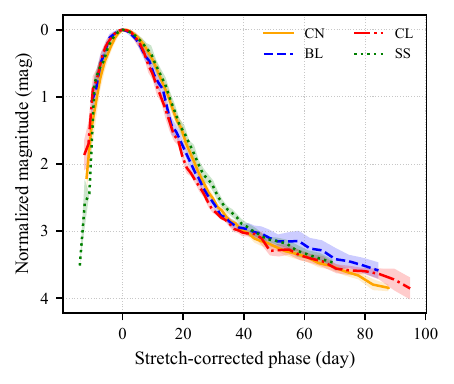}\\
    (b) $B$-band
  \end{minipage}
  \hfill
  \begin{minipage}{0.45\textwidth}
    \centering
    \includegraphics[width=8.0cm]{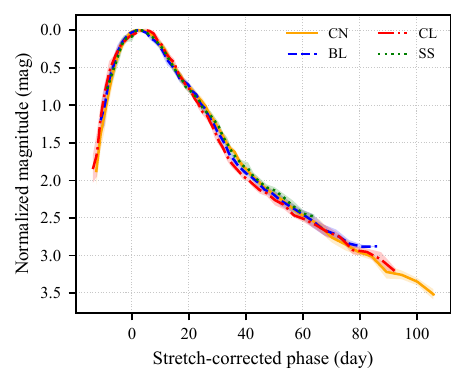}\\
    (c) $V$-band
  \end{minipage}
  \hfill
  \begin{minipage}{0.45\textwidth}
    \centering
    \includegraphics[width=8.0cm]{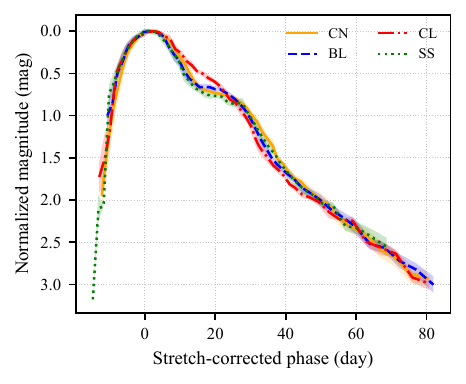}\\
    (d) $R$-band
  \end{minipage}
  \begin{minipage}{0.45\textwidth}
    \centering
    \includegraphics[width=8.0cm]{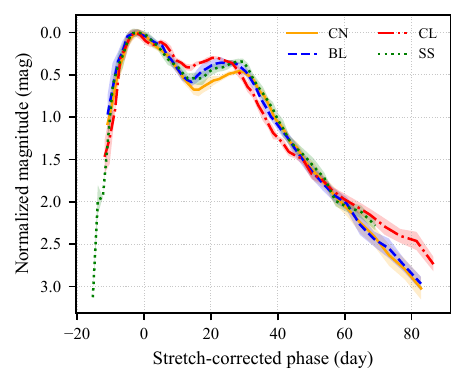}\\
    (e) $I$-band
  \end{minipage}
  \vspace{0.6em}
  \caption{Average light curves of each Branch subtype in the $U$, $B$, $V$, $R$, and $I$-bands. Epochs are in stretch-corrected phases relative to the $B$-band maximum. Peak magnitudes are normalized to 0.0\,mag. Colors and linestyles correspond to Branch subtypes: CN (orange solid), BL (blue dashed), CL (red dash-dotted), and SS (green dotted). Shaded regions indicate 1$\sigma$ dispersions of the sample in each bin. {Alt text: The figure consists of five panels corresponding to the UBVRI bands. Each panel shows average light curves of the four Branch subgroups, with stretch-corrected phase on the x axis and normalized magnitude on the y axis.}}
  \label{fig:average_lc}
\end{figure*}

Figure~\ref{fig:average_lc} shows the resulting average light curves for each Branch subtype across the $U$, $B$, $V$, $R$, and $I$ bands. In these plots, the peak magnitudes have been normalized to 0.0 mag to facilitate comparison between subtypes. To illustrate the data coverage and dispersion, all individual light curves used to construct these averages are overplotted in appendix~\ref{app:individual_lc} (figure~\ref{fig:individual_lc}).

\begin{figure}[htbp]
  \centering
  \includegraphics[width=8.0cm]{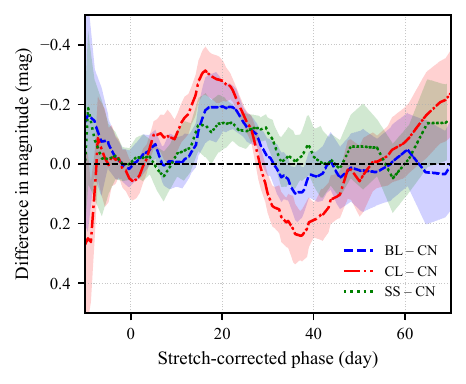}
  \caption{
  Residuals of the $I$-band average light curves of BL, CL, and SS subtypes relative to CN, shown as a function of stretch-corrected phase. Shaded regions indicate the $1\sigma$ uncertainties propagated from the dispersions of the two subtypes.
  {Alt text: The x axis shows the stretch-corrected phase, and the y axis shows the difference in magnitude between core normal supernovae and those from the other groups.}}
  \label{fig:residual_cn}
\end{figure}

Notably, in the $I$-band, the BL, CL, and SS light curves exhibit a more prominent secondary maximum relative to the primary peak compared to the CN light curve (figure~\ref{fig:residual_cn}). The pronounced secondary maximum seen in SS objects is consistent with theoretical predictions that more luminous SNe Ia exhibit stronger secondary maxima (e.g., \cite{kasen06}). The trend in CL is noteworthy given its departure from the conventional view of CL (91bg-like) events, which are generally understood to display weak or absent secondary peaks in the NIR. Note that CL SNe Ia with no secondary peaks have been excluded from the sample for making average light curves because of their poor fits; these results are solely from CL SNe Ia that have secondary peaks. 

The differences between CN and BL are newly found as well. While previous work (e.g., \cite{burgaz21}) identified an excess in the I-band around the secondary peak in HV SNe Ia, this finding has not been broken down by the Branch subclass, and by a sample size of this study. Thus, our findings that BL SNe Ia (which often overlap with the HV category) show enhanced I-band secondary maxima represents a novel contribution to the characterization of subclass diversity.

Interestingly, we also find that the CL light curve shows a more gradual decline after about 50 days past $B$-band maximum. While previous studies have emphasized their rapid post-maximum evolution (e.g. \cite{taubenberger08}), the relative shallowness of their late-time decline has not, to our knowledge, been reported before. In the R-band, the CL subgroup also deviates from the other groups by displaying a slower decline after the first peak and an almost absent plateau.

\subsection{Color curves} \label{color_curve}
\begin{figure}[ht]
    \centering
    \includegraphics[width=8.0cm]{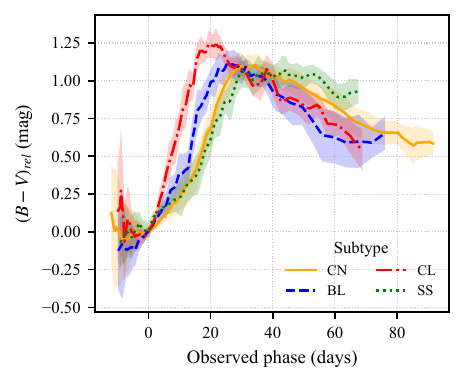}
    \caption{$(B-V)_{\rm rel}$ color curves of SNe~Ia, grouped by Branch subtype, shown as a function of observed phase relative to $B$-band maximum. Curves are normalized to $(B-V)=0$ at maximum light. {Alt text: A line graph with the x axis as observed phase, and the y-axis as the color relative to B-band maximum.}}
    \label{fig:bv_curve}
\end{figure}

The evolution of SN colors reflects underlying physical properties, such as the temperature of the ejecta and the reprocessing of blue light into longer wavelengths by iron-group element lines. To study this behavior, we used the average light curves to derive average $B-V$ color curves for each subtype (figure \ref{fig:bv_curve}). In order to compute $B-V$ consistently, we placed the mean $B$ and $V$-band light curves on the observed phase using their corresponding average stretch factors. This procedure placed the two bands on the same observed time axis, ensuring that their relative timescales were accurately reproduced. The color curves have been normalized so $B-V=0$ at B-band maximum; hereafter, we denote color as $(B-V)_{\rm rel}$, the value relative to the peak color.

All subtypes exhibit an initial reddening phase up to $\sim$20-30 days past $B$-band maximum, followed by a gradual return toward bluer colors. Notably, the CL subtype reaches the maximum peak color ($\sim$ 1.18 mag) most rapidly, while the SS subtype evolves more slowly and peaks at a slightly lower $(B-V)_{\rm rel}$ value ($\sim$ 1.08 mag). CN and BL have similar peak colors to SS, but their timings vary. These differences suggest systematic variations in fluorescence among the subtypes. In table \ref{tab:color_peak} we summarize the peak $(B-V)_{\rm rel}$ values and epochs for each subtype, estimated by smoothing the color curves.

\begin{table}[htbp]
  \tbl{Peak epochs and amplitudes of the $B-V$ color curves for each Branch subtype. Values were measured from Savitzky-Golay-smoothed profiles (window length$=51$ days, polynomial order$=3$).}{%
  \begin{tabular}{lcc}
      \hline
      Subtype & Epoch of $(B-V)_{\rm rel,max}$ & $(B-V)_{\rm rel,max}$ \\ 
      \hline
      CN  & $32.2\pm1.9$ & $1.097\pm0.034$ \\
      BL  & $26.1\pm1.4$ & $1.106\pm0.029$ \\
      CL  & $19.5\pm1.3$ & $1.239\pm0.040$ \\
      SS  & $43.9\pm6.2$ & $1.054\pm0.070$ \\
      \hline
    \end{tabular}}\label{tab:color_peak}
\begin{tabnote}
\end{tabnote}
\end{table}

It is worth noting that the SS color curve evolves much more slowly after maximum, nearly forming a plateau, whereas the other subgroups become bluer at broadly similar rates. Between phases $+30$ and $+90$ d, the $B-V$ colors of SNe Ia is known to be approximately linear (the Lira relation; \cite{lira95}; \cite{phillips99}). More recent work indicates that the Lira-phase slope depends on light-curve shape (e.g., color stretch $s_{BV}$; \cite{burns14}). These results may expand the known diversity in $B-V$ evolution, with SS occupying the shallow slope end of the distribution.

Overall, the ordering of peak color timing largely follows the known stretch/$\Delta m_{15}(B)$ sequence: CL events redden and peak earliest, BL and CN show intermediate behavior, and SS evolves most slowly. This consistency suggests that the diversity in color evolution is closely linked to the width–luminosity relation and ultimately to the amount of synthesized $^{56}$Ni and its distribution.

\subsection{SNe Ia with poor fits} \label{ssec:poor_fit}
We define "poor fit SNe Ia" as objects with a reduced chi-square value of $\chi_{\rm red}^2 \geq 4$ in the first iteration when fitted to the initial template. The number of poor fit samples in each Branch subgroup is presented in table~\ref{tab:n_branch}.

\begin{figure}[ht]
    \centering
    \includegraphics[width=8.0cm]{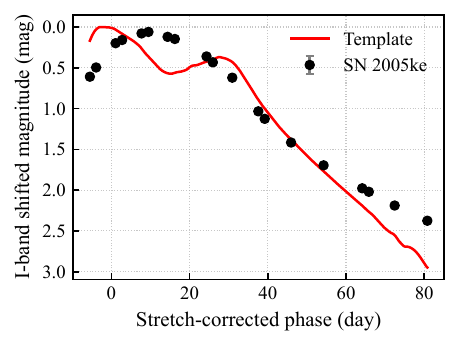}
    \caption{The fit $I$-band light curve of CL SN Ia SN 2005ke (black circles) with the initial TAK08 template (red solid curve). Most of the poor fit SNe Ia have peculiar $I$-band light curves with only one peak like this SN, resulting in $\chi_{\rm red}^2\geq4$. {Alt text: A graph. The x axis shows stretch-corrected phase, and the y axis shows $I$-band shifted magnitude.}}
    \label{fig:poorfit_example}
\end{figure}

Most of the poor fits arise from the absence of a clear secondary maximum in the $I$-band light curve, the lack of a plateau in the $R$-band light curve, or both (an example is shown in figure~\ref{fig:poorfit_example}). This behavior is particularly evident in the CL subgroup. In these subluminous events, the ejecta cools more rapidly after maximum, and the photospheric temperature reaches $\sim7,000$ K earlier than normal SNe Ia. At this temperature, the secondary maximum would usually appear, but in CL events it occurs so early that it merges with the primary peak. The resulting overlap of the first and second peaks causes deviations from the expected template shape and leads to higher $\chi_{\rm red}^2$ values. In addition, some SNe Ia exhibit poor fits due to insufficient data around the peak or in the $R$/$I$ bands, where the light curve morphology is more complex than in other bands.

\section{Discussion}
The pronounced diversity of $I$-band light curves among SNe Ia, especially in the secondary maximum and the late-time decline, provides a powerful diagnostic of explosion physics. Here, we first demonstrate that the Fe II line evolution, rather than continuum or Ca II features, drives the diversity of the secondary maximum. We then show how the late-time $I$-band slope can serve as a photometric indicator of the CL subclass.

\subsection{Diversity of the $I$-band secondary maximum}
Theoretical work attributes the $I$-band secondary maximum to iron recombination as the ejecta cools to $\sim7,000$ K, causing Fe II lines to emerge (\cite{kasen06}; \cite{jack15}). To test this idea, we took the CSP spectrum data in the $i$-band wavelength window (6,960-8,340 \AA, corresponding to $\geq40\%$ of the maximum transmittance) and divided them into three regions: (1) a Fe II region (7,300-8,000 \AA), (2) a non-Fe II region including the adjacent continuum plus Ca II absorption (6,960-7,300 \AA~and 8,000-8,340 \AA), and (3) a complementary redward Ca II emission region (8,340-8,870 \AA). After normalizing each spectrum to its interpolated photometric magnitude, we generated synthetic light curves by integrating flux within these regions. We do not use spectra from the TAK08 sample, as they do not cover the full $I$-band wavelength range.

\begin{figure}[htbp]
    \centering
    \includegraphics[width=8.0cm]{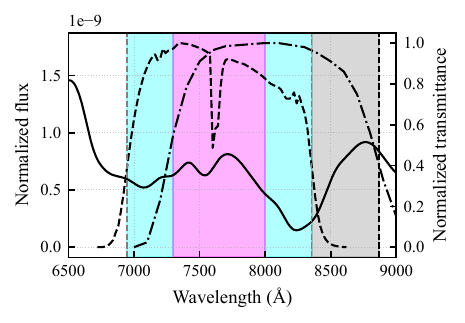}\\
    \includegraphics[width=8.0cm]{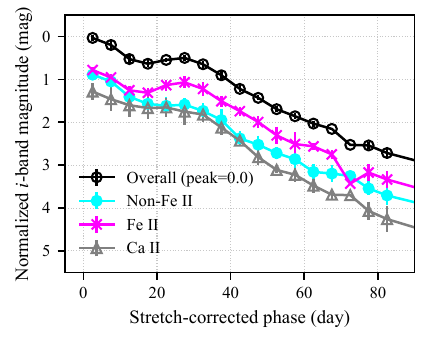}\\
    \caption{The top panel shows the wavelength ranges used for the decomposition of the $i/I$-band. The black solid line shows the Hsiao spectral template at +28 days, while the black dashed and dot–dashed lines denote the CSP $i$ and Bessell $I$ filter transmission curves, respectively. The magenta band marks the Fe II region (7,300–8,000 \AA), the gray band indicates the Ca II region (8,340-8,870 \AA), and the cyan bands correspond to the non–Fe II contributions (6,960-7,300 \AA~and 8,000-8,340 \AA). Vertical dashed lines mark the 40\% transmittance thresholds of each filter. The bottom panel presents the corresponding synthetic light curves in stretch-corrected phase, normalized to zero at peak. Shown are the overall light curve (black open circles), the non–Fe II contribution (cyan filled circles), the Fe II contribution (magenta crosses), and the Ca II contribution (gray open triangles). {Alt text: In the top panel, the x axis shows wavelength. The y axis is used for two quantities, normalized flux and transmittance. In the bottom panel, the x axis shows stretch-corrected phase, and the y axis shows the normalized i-band magnitude.}}
    \label{fig:regional_lc}
\end{figure}

\begin{figure}[ht]
    \centering
    \includegraphics[width=8.0cm]{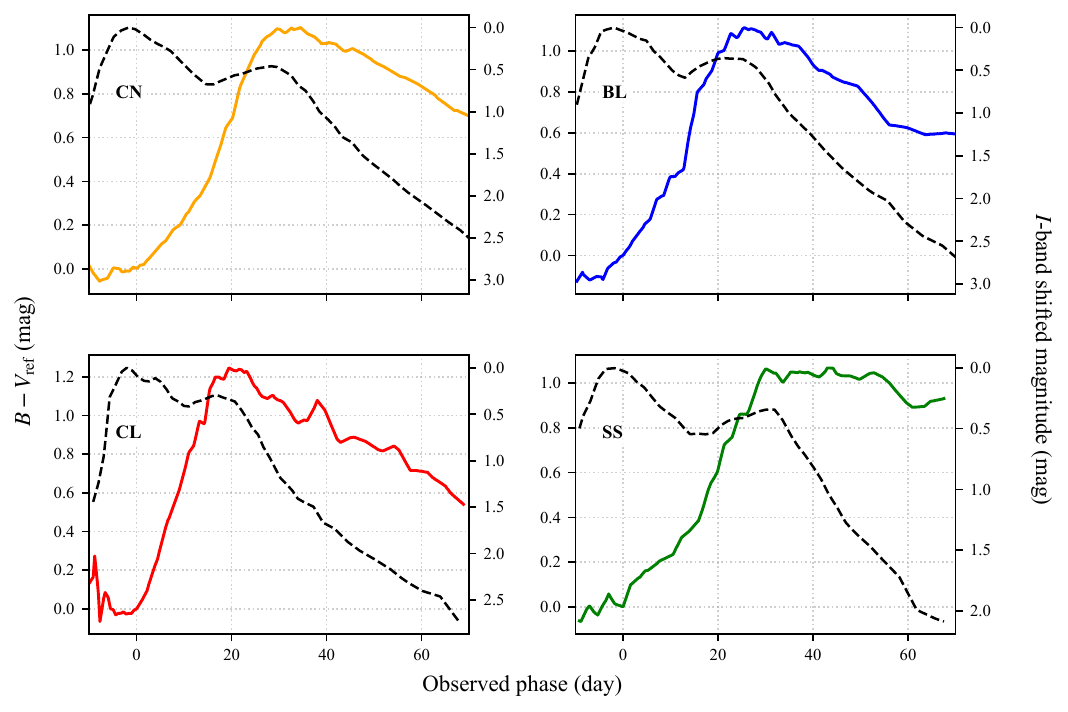}
    \caption{Normalized synthetic light curves for each Branch subtype, comparing the overall $I$-band magnitude (black dashed line) to the $(B-V)_{\rm rel}$ color evolution (solid colored line) on a common observed phase axis. {Alt text: The figure consists of four panels, each corresponding to different Branch subgroups. In each panel, the x axis shows the observed phase. The y axis is used for two quantities, the B-V colors relatived to B-band maximum, and shifted i-band magnitudes.}}
    \label{fig:color_2peak}
\end{figure}

Figure \ref{fig:regional_lc} presents these wavelength divisions alongside their corresponding synthetic light curves. The results show that the Fe II region dominates the secondary maximum, whereas the non-Fe II and Ca II region does not show a strong bump; consequently, Ca II emission features are excluded from further discussion of subclass-dependent spectral diversity. Moreover, the observation that all subclasses exhibit similar color at the time of the secondary peak (figure \ref{fig:color_2peak}), implies that the emergence of Fe II features occurs under comparable thermal conditions, providing statistical support for models that link the $I$-band secondary maximum to recombination of iron‐group elements. This further indicates that the recession of the photosphere is coupled with the iron recombination front, linking the observed color evolution directly to the transition from photosphere-powered to line-powered luminosity. This further indicates that the recession of the photosphere is coupled with the iron recombination front. These results are consistent with the theoretical prediction of \citet{kasen06}, and the spectroscopic evidence found by \citet{jack15} and \citet{deckers25}, all of which point to Fe II emission as the dominant origin of the secondary maximum.

\begin{figure}[ht]
    \centering
    \includegraphics[width=8.0cm]{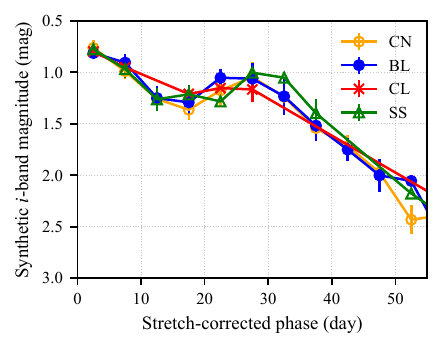}
    \caption{Subtype‐resolved synthetic light curves in stretched-corrected phases computed solely within the Fe II region (7,150–8,000 \AA). {Alt text: A line graph. The x axis shows stretch-corrected phase, and the y axis shows normalized i-band magnitude.}}
    \label{fig:feii_lc_subtype}
\end{figure}

Subtype-specific Fe II light curves are presented in figure \ref{fig:feii_lc_subtype}. Note that these light curves (as well as the light curves in figure \ref{fig:regional_lc}) have been corrected for stretch. The diversity seen in this figure is slightly different from the overall $I$-band light curves (figure \ref{fig:average_lc}); here, the amplitudes of the secondary maximum of CN, BL, and SS do not change greatly. This suggests that although the non-Fe II regions do not contribute to the secondary maximum significantly, changes in its light curves like the slope could change the strength of the bump.

Hereafter, we focus on the diversity of the Fe II region light curves, which had the greatest contribution to the secondary maximum. The amplitude still shows some variety, with CL having a weaker secondary maximum than the other subgroups. However, it is rather the timing that varies for different subgroups, with CL and BL having earlier peaks than CN and SS. The significance of this trend can be further confirmed by collecting more spectra around the secondary maximum, because figure \ref{fig:feii_lc_subtype} is binned by 5 days due to the limited number of spectra.

\subsection{Interpretations from theoretical models}
\citet{kasen06} performed radiative transfer simulations for certain ejecta structures to calculate the effects of different physical parameters to the behavior of the secondary maximum in NIR bands, including the I-band. The parameters discussed were the degree of ${}^{56}$Ni mixing, ${}^{56}$Ni mass, and the amount of stable iron elements. Here, we discuss each parameter and its possible effect on our results.

Variation in ${}^{56}$Ni mass primarily affects the peak luminosity and the overall time scale of the light curve. Observationally, \citet{folatelli13} reported that the mean peak absolute magnitudes follow the sequence CL < BL < CN < SS. Since the luminosity of SNe Ia is proportional to the amount of ${}^{56}$Ni, this suggests that the ${}^{56}$Ni mass systematically increases from CL to SS. The overall timescale, represented by the stretch parameter, also follows the same order (CL < BL < CN < SS), consistent with the interpretation from peak luminosity diversity. However, as mentioned in section \ref{ssec:fit_result}, the stretch parameter is not solely determined by ${}^{56}$Ni mass, as it also depends on the total ejecta mass through the diffusion timescale (\cite{arnett82}). Therefore, while the correlation between peak luminosity and stretch implies a major influence from ${}^{56}$Ni, the role of the ejecta mass cannot be ignored. The effects of these two masses can be distinguished in future work by directly measuring ${}^{56}$Ni mass from nebular lines (\cite{childress15, mazzali15}). A larger ${}^{56}$Ni mass tends to delay the secondary maximum and enhance its strength, which could in principle account for part of the observed diversity.

In this study, we use stretch-normalized light curves in our discussion (figures~\ref{fig:average_lc} and \ref{fig:feii_lc_subtype}), which is expected to largely remove the first-order effect of ${}^{56}$Ni mass. If the timing of the secondary maximum were insensitive to ejecta mass, such normalization could mix ejecta mass diversity into the stretch-corrected phase. However, several observational studies suggest that the secondary maximum is related to the ejecta mass. \citet{papadogiannakis19} found that SNe~Ia with longer transparency timescales, which serve as an observational proxy for larger ejecta masses, tend to show later secondary maxima. Furthermore, \citet{dhawan15} reported that the timing of the secondary maximum in the near-infrared correlates with the $B$-band decline rate (i.e. stretch), which reflects variations in both ${}^{56}$Ni and ejecta mass. Since $B$-band light curves become strikingly uniform after stretch correction, it is likely that this procedure effectively removes the diversity arising from both ${}^{56}$Ni and ejecta masses. Based on these considerations, we assume that the stretch normalization approximately compensates for the combined effects of ${}^{56}$Ni and ejecta mass for the $I$-band, and we proceed with this interpretation in the following discussion.

The residual diversity after correcting for stretch manifests almost exclusively in the timing of the secondary maximum, and cannot be explained by ${}^{56}$Ni mass and ejecta mass alone. This indicates that an additional parameter must be invoked that alters the timing without significantly changing the amplitude. Nevertheless, the influence of ${}^{56}$Ni mass and ejecta mass cannot be regarded as completely absorbed by stretch: for instance, brighter SS events may retain higher temperatures that could delay the onset of recombination. Although the stretch-corrected color curves are broadly consistent across subtypes, implying similar temperature evolutions, we do not entirely rule out a residual effect of these two masses.

One leading candidate that can explain the remaining diversity is the mass of stable iron group elements (IGEs) synthesized in the explosion. Stable IGEs do not contribute to the overall luminosity of the supernova, but they change the radius of the iron-rich core, hastening or delaying the secondary maximum. In \citet{kasen06}, the amount of stable IGEs synthesized by electron capture has an effect of changing the timing of the secondary maximum without changing the brightness, which is the trend seen in our results. In figure \ref{fig:feii_lc_subtype},  the BL and CL subgroups have earlier secondary peaks than the CN and SS subgroups, which would imply that more stable IGEs relative to $^{56}$Ni were synthesized during the explosion in the former groups. Stable IGEs can also be synthesized when the metallicity of the WD is high, but this has an effect of changing the secondary maximum amplitude as well, which is not consistent with our results. Nevertheless, analysis with NIR light curves may be able to distinguish these parameters in future work.

Mixing of $^{56}$Ni is also capable of changing the timing and brightness of the secondary maximum. However, in the case of the $I$-band, the effect shifts the primary peak along with the secondary peak, weakening the relative variations (see figure 9 of \cite{kasen06}). Therefore, we concluded that stable IGEs is likely to have a stronger effect. However, it is worth noting that when assuming a completely homogenized composition structure, the first and second peaks become blended together to form a single late peak, which is consistent with some CL SNe Ia which did not fit well to the template. Mixing cannot be completely neglected, and other diagnostics are required to robustly distinguish between the two effects in future work. One candidate could be the shape of the early light curves, as it is predicted that SNe Ia with different ${}^{56}$Ni distribution have different early light curve behavior (\cite{magee18}).

\citet{kasen06} also found that the Ca II line can affect the secondary maximum in the I-band, depending on assumptions of radiative transfer: pure scattering and pure absorption. It was concluded that if calcium lines were from pure scattering, it would have only a modest effect on the secondary maximum. On the other hand, if pure absorption was assumed, the luminosity would change substantially. In this study, we have shown that the amplitude of the secondary maximum does not vary drastically (especially after the effect of $^{56}$Ni mass is corrected for), and that Ca II does not have a significant contribution to the bump (figure \ref{fig:regional_lc}), so this effect can be neglected. Observations are more consistent with the pure scattering case. 

\subsection{Physical differences between the Branch subgroups}
Previous studies have suggested that the CL, CN, and SS subtypes may form a continuous sequence, broadly consistent with the stretch-magnitude relation and largely driven by differences in the amount of synthesized ${}^{56}$Ni (\cite{nugent95}; \cite{branch06}).

Our results indicate that the timing of the secondary maximum (after correcting for stretch) is particularly sensitive to the amount of stable IGEs, making this parameter a strong candidate for driving the observed diversity. CL and BL events show earlier secondary maxima than CN and SS, implying that they produced larger amounts of stable IGEs. Given that the ${}^{56}$Ni mass increases in the order of CL, BL, CN, SS, we see an anti-correlation between stable IGE mass and $^{56}$Ni mass; subgroups that synthesize more $^{56}$Ni tend to produce fewer stable IGEs.

This trend is broadly consistent with the near-$M_{\rm Ch}$ delayed-detonation scenario, where high-luminosity explosions with more $^{56}$Ni tend to produce less IGEs due to less efficient electron capture (\cite{krueger10}; \cite{seitenzahl13}). On the other hand, the sub-$M_{\rm Ch}$ double-detonation scenario predicts the opposite trend: both $^{56}$Ni and stable IGEs (e.g., $^{58}$Ni, $^{57}$Fe) increase with the progenitor mass, leading to a positive correlation between the yields of these elements (\cite{woosley11}). Therefore, the observed sequence from CL to SS is naturally explained within the framework of the near-$M_{\rm Ch}$ delayed detonation scenario, while it is inconsistent with the expectations of the sub-$M_{\rm ch}$ double-detonation scenario.

Within this delayed-detonation sequence, BL events may represent an intermediate case, with the amount of stable IGE between CL and CN. However, some studies have interpreted the spectroscopic differences between BL and CN as reflecting differences in ejecta structure, such as viewing angles in an asymmetric explosion (\cite{maeda10a}) or outward mixing of intermediate mass and iron group elements (\cite{mazzali05}; \cite{stehle05}). In such scenarios, the relatively early secondary maximum could arise from a shallower distribution of radioactive material in the ejecta. BL events might even originate from different explosion mechanisms altogether, such as the sub-$M_{\rm Ch}$ double-detonation scenario, rather than representing an intermediate state within a continuous near-$M_{\rm Ch}$ sequence. This would provide an alternative explanation for the smaller stretch observed in BL relative to CN. Comparison of $I$-band light curves from different explosion models could further develop this discussion.

Taken together, our results provide new observational evidence that the diversity of the $I$-band secondary maximum is best explained within a near-$M_{\rm Ch}$ delayed-detonation framework, which naturally accounts for the observationally inferred anti-correlation between ${}^{56}$Ni and stable IGEs. While alternative interpretations (e.g., the sub-$M_{\rm Ch}$ double-detonation) cannot be fully excluded, the predicted positive correlation in those models is inconsistent with our data, except possibly for the BL subclass where additional factors such as asymmetry or mixing may play a role.

\subsection{Diversity of the late-time $I$-band tail slope}
\begin{figure}[ht]
    \centering
    \includegraphics[width=8.0cm]{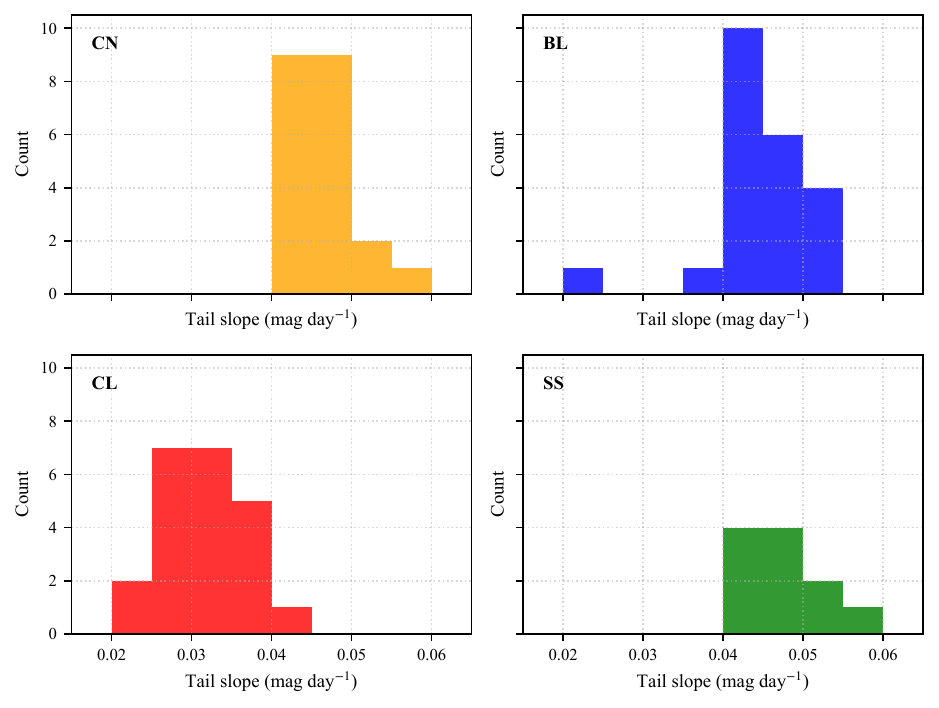}
    \caption{Normalized histogram of the $I$-band tail slope measured at $t_{\rm B,max}\ge40,$days, again color-coded by subclass. The CL events (red) occupy a distinct, more gradual locus compared to BL (blue), CN (orange), and SS (green). {Alt text: An ensemble of four panels, corresponding to each Branch subtype. Each panel is a histogram. The x axis shows the tail slope in units of magnitude per day. The y axis shows the number of supernovae.}}
    \label{fig:tail_slope}
\end{figure}

To further search for photometric indicators that distinguish the Branch subtypes, we examined the late-time decline in units of magnitude per day by fitting a straight line to data at $t_{B,{\rm max}}\ge40$ days, again discarding any SNe with visibly poor fits. Figure \ref{fig:tail_slope} illustrates the slope distribution, where CL objects occupy a distinct locus with systematically gradual declines compared to BL, CN, or SS. This clear separation implies a practical photometric criterion; by measuring the I-band slope from just two observations beyond 40 days after B-band maximum, one can efficiently identify CL type events.

\begin{figure}[ht]
    \centering
    \includegraphics[width=8.0cm]{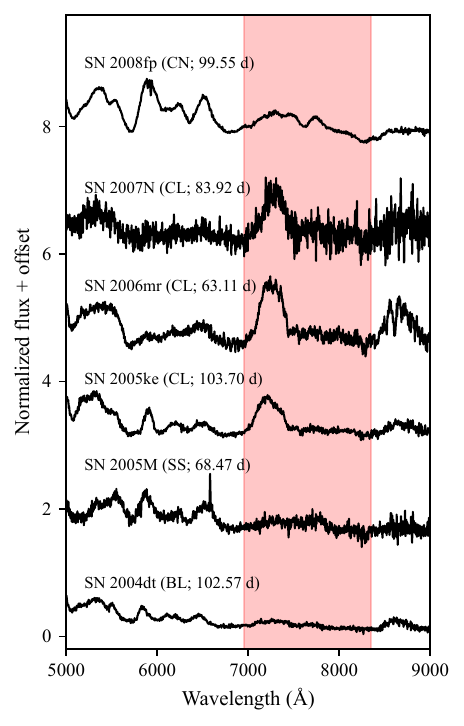}
    \caption{Late-phase spectra of three CL objects (SN 2007N, SN 2006mr, and SN 2005ke), a CN object (SN 2008fp), a BL object (SN 2004dt), and a SS object (SN 2005M). The red region shows the wavelength range in which the transmittance of the CSP $i$-band is greater than 40\% of the maximum value. Data are taken from \citet{folatelli13}. {Alt text: A graph showing six spectra. The x axis shows wavelength, and the y axis shows normalized flux with an offset for visualization.}}
    \label{fig:emission_cl}
\end{figure}

We cannot conclude exactly what causes the gradual slope seen in CL events, because $I$-band spectra of the SNe Ia used for the CL average light curves at $t_{B,{\rm max}}>50$d were unavailable for this sample. However, it is likely to be from emission features, because in this phase the ejecta has already become optically thin, allowing the contribution of emission lines to the luminosity to be dominant. In three CL SNe Ia that were not used in the average light curves, a strong emission feature at $\sim$7,200 \AA~is seen and may have been present in other SNe Ia in the same subgroup (figure \ref{fig:emission_cl}). Outside of our sample, \citet{srivastav17} found a similar feature in the $+93.9$ d spectrum of SN 2015bp, which is a CL SN Ia with a secondary maximum, likely falling within the sample consisting the CL average light curve. Future studies could statistically confirm this unique characteristic of CL SNe Ia. Identifying this line is beyond the scope of this paper; however, it is worth noting that similar lines are seen in 1991bg-like SNe Ia (\cite{filippenko92}; \cite{mazzali97}). In these studies, the feature is thought to be a blend of [Fe II], [Ni II], and [Ca II] lines. Especially for calcium, the existence of the Ca II line at $\sim8,500$ \AA~in the spectrum of SN 2006mr implies that the [Ca II] line is at least a portion of this feature. In the NIR, [Ni II] lines of subluminous SNe Ia have been detected as early as $\sim50$ days after B-band maximum (\cite{kumar25}). If an optical counterpart of [Ni II] can be identified in the spectrum, given that the phase is well over the half-life of ${}^{56}$Ni, it would imply that stable nickel is synthesized in a majority of CL SNe Ia, favoring near-$M_{\rm Ch}$ progenitors with a high central density, which would be consistent with our results.

\section{Conclusion}
In this study, we have constructed mean $UBVRI$ light curve templates for each Branch spectroscopic subtype (CN, BL, CL, and SS) using a sample of 109 low-redshift SNe Ia. By fitting individual light curves with a multi-band stretch method and iteratively averaging the fitted light curves, we have investigated photometric diversity from pre-maximum epochs through the late tail.

We summarize our findings as below.

\begin{itemize}
    \item
    CL SNe Ia have lower stretch values and SS SNe Ia have higher stretch values across all bands ($\overline{s}_{B,{\rm CL}}=0.802\pm0.073$, $\overline{s}_{B,{\rm SS}}=1.146\pm0.063$). Stretch distribution of CN and BL overlap, but CN have a larger average value than BL in $BVRI$ bands (e.g. $\overline{s}_{B,{\rm CN}}=1.012\pm0.091$, $\overline{s}_{B,{\rm BL}}=0.951\pm0.076$). These stretch differences likely reflect systematic variations in the ${}^{56}$Ni mass synthesized, possibly together with changes in the ejecta mass, with CL producing less and SS producing more ${}^{56}$Ni.

    \item 
    The $I$-band exhibits the most pronounced subtype-dependent differences. In particular, the behavior of the secondary maxima varies systematically, and the CL subgroup displays a notably slower decline in the $I$-band tail beyond $+40$ days.

    \item 
    The maximum of the $B-V$ color curve coincides with the secondary maximum for all subtypes, marking the transition from photosphere-powered to line-powered luminosity as the iron recombination front recedes inward.

    \item 
    Synthetic light curves demonstrate that Fe II recombination dominates the $I$-band secondary maximum. Combined with theoretical work, the diversity of the observed timing of the secondary maximum is best explained by the combination of $^{56}$Ni mass and the amount of stable IGEs. Earlier stretch-corrected peaks in CL and BL imply larger stable-iron cores, consistent with the anti-correlation between stable IGEs and ${}^{56}$Ni expected in the near-$M_{\rm Ch}$ delayed detonation scenario, and inconsistent with the positive correlation predicted by the sub-$M_{\rm Ch}$ double-detonation scenario. Mixing effects may also contribute and should be tested in future work.

    \item 
    The late-time $I$-band tail slope for CL SNe Ia is significantly shallower than the other subgroups. This suggests that CL SNe Ia can be distinguished from other subgroups from only two photometric observations after 40 days past $B$-band maximum. Luminosity at this phase is powered by emission lines, so a feature in the $I$-band is expected to cause this difference. However, we cannot confirm with this dataset because there was no available spectrum of a CL SN Ia that was well fit to the initial template at these epochs. Inspection of poorly fit SNe Ia (SNe Ia with only one peak in the $I$-band) showed a strong emission feature at $\sim7,200$\AA, likely a blend of [Fe II], [Ni II], and [Ca II] lines. Future work may confirm if this feature is seen in other CL events as well.
\end{itemize}

Our findings thus connect post-peak photometric behavior to explosion physics. In future work, using samples from large surveys such as the Nearby Supernova Factory (\cite{aldering02}) and the Zwicky Transient Facility (\cite{bellm18}; \cite{dhawan22}; \cite{rigault25}) could provide even more statistically robust constraints on the differences of evolution across various subclasses. In addition, extending these approaches to NIR bands and incorporating detailed modeling of mixing and electron capture effects will further disentangle the complex diversity seen in SNe Ia.

\begin{ack}
We thank the anonymous referee for the valuable and constructive comments, which greatly helped to improve the clarity and robustness of this work. This research has made use of the CfA Supernova Archive, which is funded in part by the National Science Foundation through grant AST 0907903.
\end{ack}

\section*{Funding}
This work was financially supported by Grants-in-Aid for Scientific Research (KAKENHI) of Japan Society for the Promotion of Science (JSPS) numbers 18H05223, 21H04491, 23H05432, 24H01810, 24K21204, and 24KK0070. This work was also supported by JSPS Bilateral Program Number JPJSBP 120227709.

\appendix %%%%%%%%%%%%%%%%%%%%%%%%%%%%%%%%%%%%%%%%%%%%%%%%%%%%%%%%
\section{Fitting results for each SN Ia}\label{app:fit_result}
Table~\ref{tab:fit_result} summarizes the light curve fitting results for the SN Ia sample.

\begin{longtable}{llcccccccc}
  \caption{Branch subtype, data sample, heliocentric redshift, best-fit stretch values, and the reduced chi-square for each individual SN Ia. Uncertainties are given in parentheses.}\label{tab:fit_result}  
\hline\noalign{\vskip3pt} 
  SN name & Subtype & Reference & $z$ & $s^{\rm fit}_U$ & $s^{\rm fit}_B$ & $s^{\rm fit}_V$ & $s^{\rm fit}_R$ & $s^{\rm fit}_I$ & $\chi^2_{\rm red}$\\   [2pt] 
\hline\noalign{\vskip3pt} 
\endfirsthead      
\hline\noalign{\vskip3pt} 
  SN name & Subtype & Reference & $z$ & $s^{\rm fit}_U$ & $s^{\rm fit}_B$ & $s^{\rm fit}_V$ & $s^{\rm fit}_R$ & $s^{\rm fit}_I$ & $\chi^2_{\rm red}$\\  [2pt] 
\hline\noalign{\vskip3pt} 
\endhead
\hline\noalign{\vskip3pt} 
\endfoot
\hline\noalign{\vskip3pt} 
\multicolumn{10}{p{\linewidth}}{\footnotesize
\textbf{Notes.}—
References are as follows: (1) \citet{wells94}, (2) \citet{lira98}, (3) \citet{richmond95}, (4) \citet{riess05}, (5) \citet{jha06}, (6) \citet{jha99}, (7) \citet{garnavich04}, (8) \citet{stritzinger02}, (9) \citet{krisciunas03}, (10) \citet{pignata04}, (11) \citet{anupama05}, and (12) \citet{krisciunas17}. SNe Ia from references (1)--(11) are treated as the TAK08 sample, and those from reference (12) are treated as the CSP sample in this study.
}\\
\endlastfoot 
SN 1989B & CL & (1) & 0.002 & 0.803 (0.016) & 0.931 (0.023) & 0.929 (0.013) & 0.929 (0.018) & 0.927 (0.042) & 0.710\\
SN 1990N & CN & (2) & 0.003 & 0.761 (0.006) & 1.071 (0.006) & 1.049 (0.009) & 1.088 (0.007) & 1.073 (0.005) & 2.384\\
SN 1991T & SS & (2) & 0.006 & 0.725 (0.023) & 1.127 (0.005) & 1.067 (0.004) & 1.073 (0.003) & 1.061 (0.004) & 3.077\\
SN 1994D & CN & (3) & 0.002 & 0.753 (0.013) & 0.889 (0.002) & 0.797 (0.005) & 0.829 (0.002) & 0.840 (0.001) & 1.603\\
SN 1994ae & CN & (4) & 0.004 & 0.928 (0.118) & 1.110 (0.010) & 1.014 (0.009) & 1.035 (0.006) & 1.058 (0.006) & 0.663\\
SN 1997br & SS & (5) & 0.007 & 0.775 (0.176) & 0.873 (0.275) & 1.067 (0.190) & 1.029 (0.133) & 0.981 (0.082) & 2.299\\
SN 1997cn & CL & (5) & 0.016 & 1.210 (0.366) & 1.184 (0.093) & 0.828 (0.101) & 0.673 (0.040) & 0.627 (0.095) & 19.246\\
SN 1997do & BL & (5) & 0.010 & 0.837 (0.039) & 1.011 (0.014) & 1.035 (0.010) & 1.059 (0.008) & 1.074 (0.005) & 0.781\\
SN 1997dt & CN & (5) & 0.007 & 0.924 (0.045) & 0.996 (0.043) & 0.996 (0.045) & 1.019 (0.019) & 0.992 (0.041) & 1.350\\
SN 1998V & CN & (5) & 0.018 & 0.852 (0.015) & 1.030 (0.011) & 0.967 (0.006) & 1.009 (0.008) & 0.994 (0.009) & 0.597\\
SN 1998ab & SS & (5) & 0.027 & 0.762 (0.124) & 0.992 (0.015) & 1.046 (0.008) & 1.086 (0.009) & 1.055 (0.007) & 2.287\\
SN 1998aq & CN & (4) & 0.004 & 0.902 (0.031) & 1.012 (0.005) & 0.932 (0.004) & 0.972 (0.005) & 1.002 (0.004) & 1.533\\
SN 1998bp & CL & (5) & 0.010 & 0.641 (0.013) & 0.699 (0.011) & 0.684 (0.007) & 0.677 (0.006) & 0.645 (0.005) & 3.359\\
SN 1998bu & CN & (6) & 0.003 & 0.906 (0.007) & 1.033 (0.004) & 0.931 (0.003) & 0.961 (0.005) & 0.970 (0.004) & 0.570\\
SN 1998dh & BL & (5) & 0.009 & 0.825 (0.020) & 0.962 (0.008) & 0.926 (0.006) & 0.943 (0.007) & 0.939 (0.005) & 0.314\\
SN 1998ec & BL & (5) & 0.020 & 0.827 (0.082) & 1.049 (0.069) & 0.915 (0.019) & 1.045 (0.026) & 1.027 (0.043) & 0.694\\
SN 1998eg & CN & (5) & 0.025 & 0.956 (0.055) & 0.997 (0.038) & 0.919 (0.029) & 0.930 (0.103) & 1.296 (0.482) & 1.163\\
SN 1998es & SS & (5) & 0.011 & 0.905 (0.021) & 1.183 (0.016) & 1.032 (0.010) & 1.084 (0.015) & 1.127 (0.012) & 0.598\\
SN 1999aa & SS & (5) & 0.014 & 0.954 (0.007) & 1.168 (0.007) & 1.070 (0.006) & 1.114 (0.007) & 1.117 (0.006) & 0.805\\
SN 1999ac & SS & (5) & 0.009 & 0.873 (0.037) & 1.052 (0.009) & 0.982 (0.003) & 1.017 (0.004) & 0.972 (0.005) & 1.648\\
SN 1999by & CL & (7) & 0.002 & 0.682 (0.060) & 0.661 (0.030) & 0.646 (0.008) & 0.656 (0.010) & 0.605 (0.005) & 12.339\\
SN 1999cc & BL & (5) & 0.031 & 0.798 (0.057) & 0.868 (0.016) & 0.827 (0.011) & 0.820 (0.010) & 1.272 (0.379) & 2.337\\
SN 1999cl & BL & (5) & 0.008 & 0.721 (0.032) & 0.996 (0.023) & 0.970 (0.030) & 0.984 (0.074) & 0.904 (0.064) & 0.691\\
SN 1999dq & SS & (5) & 0.014 & 0.900 (0.007) & 1.125 (0.005) & 1.053 (0.005) & 1.089 (0.006) & 1.109 (0.009) & 0.942\\
SN 1999ee & SS & (8) & 0.011 & 0.864 (0.002) & 1.149 (0.002) & 1.052 (0.002) & 1.067 (0.002) & 1.080 (0.002) & 0.884\\
SN 1999ej & BL & (5) & 0.014 & 0.664 (0.069) & 0.820 (0.038) & 0.797 (0.056) & 0.814 (0.419) & 0.844 (0.135) & 1.090\\
SN 1999gd & BL & (5) & 0.019 & 0.890 (0.067) & 0.966 (0.015) & 0.900 (0.009) & 0.935 (0.009) & 0.939 (0.017) & 1.474\\
SN 1999gh & BL & (5) & 0.008 & 1.000 (0.181) & 0.941 (0.060) & 0.688 (0.057) & 1.100 (0.204) & 0.616 (0.855) & 19.774\\
SN 1999gp & SS & (5) & 0.027 & 0.945 (0.017) & 1.238 (0.010) & 1.146 (0.010) & 1.205 (0.013) & 1.196 (0.018) & 1.414\\
SN 2000B & BL & (5) & 0.020 & 1.477 (0.222) & 1.157 (0.153) & 0.876 (0.062) & 0.872 (0.089) & 1.085 (0.150) & 40.265\\
SN 2000E & SS & (5) & 0.005 & 0.917 (0.006) & 1.102 (0.003) & 1.002 (0.003) & 1.066 (0.006) & 0.968 (0.012) & 1.866\\
SN 2000cn & CL & (5) & 0.023 & 0.743 (0.011) & 0.773 (0.194) & 0.756 (0.011) & 0.771 (0.015) & 0.844 (0.449) & 1.132\\
SN 2000cx & SS & (5) & 0.008 & 0.944 (0.041) & 0.942 (0.023) & 1.000 (0.000) & 1.000 (0.000) & 1.000 (0.000) & 0.155\\
SN 2000dk & CL & (5) & 0.017 & 0.625 (0.010) & 0.764 (0.005) & 0.765 (0.007) & 0.756 (0.008) & 0.775 (0.018) & 1.538\\
SN 2001el & CN & (9) & 0.004 & 0.774 (0.024) & 0.985 (0.004) & 1.016 (0.002) & 0.970 (0.005) & 1.028 (0.003) & 0.725\\
SN 2002er & BL & (10) & 0.009 & 0.765 (0.014) & 0.931 (0.008) & 0.918 (0.009) & 0.915 (0.007) & 0.896 (0.005) & 0.569\\
SN 2003du & CN & (11) & 0.006 & 0.986 (0.010) & 1.049 (0.008) & 0.987 (0.007) & 1.022 (0.005) & 1.062 (0.021) & 1.154\\
SN 2004dt & BL & (12) & 0.020 & 0.928 (0.219) & 1.003 (0.086) & 0.815 (0.024) & 0.949 (0.018) & 0.921 (0.014) & 2.722\\
SN 2004ef & BL & (12) & 0.031 & 0.793 (0.006) & 0.891 (0.002) & 0.833 (0.002) & 0.841 (0.001) & 0.813 (0.001) & 0.900\\
SN 2004eo & CL & (12) & 0.016 & 0.670 (0.035) & 0.891 (0.005) & 0.847 (0.005) & 0.836 (0.005) & 0.882 (0.003) & 2.717\\
SN 2004ey & CN & (12) & 0.016 & 0.887 (0.003) & 1.068 (0.003) & 0.986 (0.002) & 1.028 (0.002) & 1.017 (0.001) & 0.541\\
SN 2004gs & CL & (12) & 0.027 & 0.687 (0.024) & 0.772 (0.005) & 0.689 (0.006) & 0.713 (0.003) & 0.759 (0.004) & 2.851\\
SN 2004gu & SS & (12) & 0.046 & 1.044 (0.016) & 1.261 (0.007) & 1.163 (0.006) & 1.190 (0.005) & 1.169 (0.005) & 1.423\\
SN 2005M & SS & (12) & 0.025 & 0.916 (0.007) & 1.208 (0.005) & 1.107 (0.004) & 1.123 (0.003) & 1.131 (0.002) & 0.829\\
SN 2005am & BL & (12) & 0.008 & 0.710 (0.005) & 0.825 (0.003) & 0.754 (0.003) & 0.764 (0.003) & 0.739 (0.002) & 0.915\\
SN 2005bg & SS & (12) & 0.023 & 0.962 (0.009) & 1.113 (0.005) & 1.002 (0.009) & 1.039 (0.007) & 1.085 (0.007) & 1.133\\
SN 2005bl & CL & (12) & 0.024 & 0.975 (0.085) & 0.976 (0.084) & 0.651 (0.008) & 0.636 (0.011) & 1.277 (0.118) & 35.609\\
SN 2005bo & CN & (12) & 0.014 & 0.728 (0.009) & 0.913 (0.011) & 0.867 (0.017) & 0.881 (0.012) & 0.888 (0.008) & 0.239\\
SN 2005el & CN & (12) & 0.015 & 0.766 (0.008) & 0.960 (0.003) & 0.790 (0.004) & 0.830 (0.003) & 0.832 (0.001) & 0.711\\
SN 2005eq & SS & (12) & 0.029 & 0.978 (0.006) & 1.235 (0.006) & 1.123 (0.006) & 1.146 (0.007) & 1.157 (0.004) & 0.894\\
SN 2005hc & CN & (12) & 0.046 & 0.964 (0.014) & 1.156 (0.007) & 1.025 (0.004) & 1.073 (0.004) & 1.045 (0.005) & 0.677\\
SN 2005kc & CN & (12) & 0.015 & 0.915 (0.005) & 0.960 (0.005) & 0.930 (0.004) & 0.969 (0.005) & 0.937 (0.004) & 1.425\\
SN 2005ke & CL & (12) & 0.005 & 0.799 (0.047) & 0.920 (0.025) & 0.670 (0.004) & 1.116 (0.020) & 0.599 (0.018) & 22.223\\
SN 2005ki & CN & (12) & 0.019 & 0.829 (0.045) & 0.921 (0.007) & 0.805 (0.005) & 0.831 (0.003) & 0.823 (0.002) & 0.814\\
SN 2005ku & CN & (12) & 0.045 & 1.063 (0.161) & 1.047 (0.022) & 1.016 (0.023) & 0.957 (0.025) & 0.928 (0.036) & 0.959\\
SN 2005na & CN & (12) & 0.026 & 0.795 (0.009) & 1.005 (0.005) & 0.872 (0.017) & 0.991 (0.003) & 0.977 (0.003) & 0.661\\
SN 2006D & CN & (12) & 0.009 & 0.776 (0.445) & 0.882 (0.046) & 0.894 (0.502) & 0.784 (0.026) & 0.801 (0.094) & 2.858\\
SN 2006X & BL & (12) & 0.005 & 0.846 (0.005) & 1.003 (0.011) & 0.990 (0.004) & 1.002 (0.002) & 0.982 (0.001) & 3.758\\
SN 2006ax & CN & (12) & 0.017 & 0.849 (0.004) & 1.063 (0.002) & 0.993 (0.003) & 1.023 (0.002) & 1.014 (0.002) & 0.873\\
SN 2006ef & BL & (12) & 0.018 & 0.891 (0.106) & 0.909 (0.027) & 0.746 (0.014) & 0.849 (0.036) & 0.838 (0.013) & 0.710\\
SN 2006ej & BL & (12) & 0.020 & 0.836 (0.047) & 0.894 (0.030) & 0.803 (0.054) & 0.800 (0.071) & 0.822 (0.017) & 0.434\\
SN 2006eq & CL & (12) & 0.050 & 1.000 (0.104) & 0.647 (0.023) & 0.644 (0.021) & 0.640 (0.010) & 0.671 (0.016) & 3.906\\
SN 2006et & CN & (12) & 0.022 & 0.892 (0.011) & 1.145 (0.005) & 1.031 (0.005) & 1.084 (0.002) & 1.056 (0.005) & 0.645\\
SN 2006fw & CN & (12) & 0.083 & 1.225 (0.410) & 1.000 (0.042) & 0.939 (0.052) & 0.940 (0.026) & 0.989 (0.041) & 8.585\\
SN 2006gj & CL & (12) & 0.028 & 0.503 (0.065) & 0.719 (0.016) & 0.685 (0.015) & 0.700 (1.419) & 0.749 (0.985) & 2.305\\
SN 2006gt & CL & (12) & 0.045 & 0.531 (0.108) & 0.741 (0.037) & 0.618 (0.009) & 0.649 (0.004) & 0.698 (0.018) & 3.903\\
SN 2006hx & SS & (12) & 0.045 & 0.795 (0.214) & 1.085 (0.036) & 0.995 (0.011) & 0.996 (0.027) & 1.020 (0.169) & 2.937\\
SN 2006is & CN & (12) & 0.031 & 1.171 (0.032) & 1.313 (0.010) & 1.069 (0.003) & 1.144 (0.007) & 1.135 (0.005) & 2.330\\
SN 2006kf & CL & (12) & 0.021 & 0.603 (0.030) & 0.778 (0.018) & 0.708 (0.015) & 0.705 (1.061) & 0.756 (0.008) & 1.789\\
SN 2006mr & CL & (12) & 0.006 & 0.686 (0.082) & 0.870 (0.015) & 0.617 (0.333) & 1.632 (0.175) & 0.525 (0.101) & 35.597\\
SN 2006os & CL & (12) & 0.033 & 0.680 (0.035) & 0.890 (0.009) & 0.882 (0.009) & 0.855 (0.005) & 0.966 (0.015) & 1.820\\
SN 2006ot & BL & (12) & 0.053 & 1.307 (0.166) & 1.273 (0.044) & 1.139 (0.022) & 1.272 (0.036) & 1.077 (0.017) & 3.908\\
SN 2007A & CN & (12) & 0.018 & 0.841 (0.021) & 1.175 (0.017) & 0.995 (0.017) & 1.010 (0.012) & 1.046 (0.009) & 0.849\\
SN 2007N & CL & (12) & 0.013 & 0.848 (0.255) & 1.015 (0.027) & 0.630 (0.006) & 0.563 (0.037) & 1.430 (0.151) & 15.897\\
SN 2007S & SS & (12) & 0.014 & 0.975 (0.005) & 1.188 (0.011) & 1.110 (0.008) & 1.116 (0.004) & 1.128 (0.003) & 1.639\\
SN 2007af & BL & (12) & 0.005 & 0.869 (0.006) & 0.989 (0.005) & 0.920 (0.003) & 0.959 (0.002) & 0.934 (0.002) & 0.768\\
SN 2007ai & SS & (12) & 0.032 & 0.914 (0.060) & 1.224 (0.017) & 1.114 (0.015) & 1.136 (0.016) & 1.137 (0.012) & 0.852\\
SN 2007al & CL & (12) & 0.012 & 0.512 (0.021) & 0.806 (0.052) & 0.622 (0.011) & 0.552 (0.005) & 0.502 (0.475) & 18.490\\
SN 2007as & BL & (12) & 0.018 & 0.890 (0.013) & 0.958 (0.007) & 0.876 (0.003) & 0.901 (0.003) & 0.875 (0.002) & 0.543\\
SN 2007ax & CL & (12) & 0.007 & 0.914 (0.073) & 1.001 (0.089) & 0.593 (0.024) & 0.984 (1.338) & 0.492 (0.814) & 71.515\\
SN 2007ba & CL & (12) & 0.039 & 0.707 (0.016) & 0.659 (0.007) & 0.607 (0.008) & 0.691 (0.009) & 0.668 (0.011) & 6.602\\
SN 2007bc & CL & (12) & 0.021 & 0.635 (0.007) & 0.858 (0.012) & 0.815 (0.003) & 0.843 (0.003) & 0.882 (0.002) & 1.314\\
SN 2007bd & BL & (12) & 0.031 & 0.866 (0.013) & 0.960 (0.006) & 0.841 (0.007) & 0.875 (0.004) & 0.902 (0.007) & 2.118\\
SN 2007bm & CN & (12) & 0.006 & 0.776 (0.007) & 0.975 (0.003) & 0.911 (0.003) & 0.915 (0.003) & 0.904 (0.002) & 0.684\\
SN 2007ca & CN & (12) & 0.014 & 0.955 (0.014) & 1.132 (0.004) & 1.010 (0.004) & 1.054 (0.003) & 1.090 (0.005) & 1.419\\
SN 2007hj & CL & (12) & 0.014 & 0.612 (0.007) & 0.757 (0.008) & 0.679 (0.007) & 0.686 (0.006) & 0.704 (0.016) & 5.533\\
SN 2007jg & BL & (12) & 0.037 & 0.870 (0.025) & 0.983 (0.009) & 0.880 (0.011) & 0.939 (0.009) & 0.917 (0.006) & 0.974\\
SN 2007le & BL & (12) & 0.007 & 0.993 (0.016) & 1.104 (0.004) & 1.010 (0.003) & 1.042 (0.002) & 1.010 (0.001) & 0.678\\
SN 2007nq & BL & (12) & 0.045 & 0.669 (0.015) & 0.834 (0.009) & 0.740 (0.007) & 0.769 (0.005) & 0.756 (0.006) & 1.069\\
SN 2007on & CL & (12) & 0.006 & 0.760 (0.133) & 0.895 (0.134) & 0.809 (0.094) & 0.597 (0.534) & 0.668 (0.489) & 10.111\\
SN 2007ux & CL & (12) & 0.031 & 0.572 (0.012) & 0.712 (0.006) & 0.681 (0.006) & 0.667 (0.004) & 0.694 (0.005) & 4.345\\
SN 2008C & SS & (12) & 0.017 & 0.875 (0.026) & 1.012 (0.014) & 1.000 (0.010) & 0.914 (0.009) & 1.021 (0.010) & 2.874\\
SN 2008R & CL & (12) & 0.013 & 0.610 (0.009) & 0.716 (0.004) & 0.650 (0.003) & 0.671 (0.125) & 0.659 (0.006) & 2.648\\
SN 2008ar & CN & (12) & 0.026 & 0.552 (0.162) & 0.987 (0.023) & 0.964 (0.005) & 0.983 (0.007) & 0.972 (0.007) & 3.830\\
SN 2008bc & CN & (12) & 0.015 & 0.961 (0.004) & 1.135 (0.002) & 0.994 (0.003) & 1.021 (0.003) & 1.028 (0.002) & 0.797\\
SN 2008bf & CN & (12) & 0.024 & 0.921 (0.008) & 1.137 (0.003) & 0.987 (0.002) & 1.037 (0.003) & 1.037 (0.004) & 0.830\\
SN 2008bq & CN & (12) & 0.034 & 0.896 (0.013) & 1.113 (0.007) & 1.037 (0.006) & 1.039 (0.004) & 1.027 (0.006) & 0.396\\
SN 2008fp & CN & (12) & 0.006 & 0.892 (0.008) & 1.137 (0.008) & 1.008 (0.004) & 1.044 (0.002) & 1.005 (0.002) & 0.728\\
SN 2008gl & BL & (12) & 0.034 & 0.804 (0.012) & 0.916 (0.004) & 0.824 (0.005) & 0.844 (0.004) & 0.829 (0.004) & 0.724\\
SN 2008hu & BL & (12) & 0.050 & 0.720 (0.016) & 0.853 (0.010) & 0.758 (0.006) & 0.766 (0.005) & 0.776 (0.005) & 1.836\\
SN 2008hv & CN & (12) & 0.013 & 0.851 (0.012) & 0.939 (0.003) & 0.838 (0.003) & 0.851 (0.002) & 0.853 (0.002) & 1.111\\
SN 2008ia & BL & (12) & 0.022 & 0.824 (0.009) & 0.949 (0.004) & 0.830 (0.004) & 0.844 (0.003) & 0.828 (0.004) & 0.633\\
SN 2009D & CN & (12) & 0.025 & 0.902 (0.004) & 1.127 (0.004) & 1.039 (0.003) & 1.097 (0.003) & 1.057 (0.004) & 0.487\\
SN 2009F & CL & (12) & 0.013 & 0.576 (0.020) & 0.634 (0.014) & 0.576 (0.011) & 0.589 (0.128) & 0.545 (0.027) & 14.607\\
SN 2009Y & BL & (12) & 0.009 & 0.948 (0.017) & 1.084 (0.010) & 1.015 (0.003) & 1.053 (0.002) & 0.996 (0.002) & 2.066\\
SN 2009aa & CN & (12) & 0.027 & 0.820 (0.005) & 0.968 (0.005) & 0.896 (0.004) & 0.928 (0.003) & 0.926 (0.003) & 1.138\\
SN 2009ab & CN & (12) & 0.011 & 0.823 (0.096) & 0.962 (0.013) & 0.889 (0.009) & 0.915 (0.009) & 0.909 (0.080) & 1.439\\
SN 2009ad & SS & (12) & 0.028 & 0.811 (0.010) & 1.085 (0.005) & 0.970 (0.004) & 0.971 (0.005) & 1.044 (0.004) & 1.151\\
SN 2009ag & BL & (12) & 0.009 & 0.881 (0.005) & 1.032 (0.003) & 0.981 (0.003) & 1.009 (0.002) & 0.998 (0.002) & 1.355\\
\end{longtable}
\vspace{1ex}

\section{Correlation between stretch and pEW parameters}\label{app:stretch_vs_pew}
\begin{figure}[htbp]
    \centering
    \includegraphics[width=6.0cm]{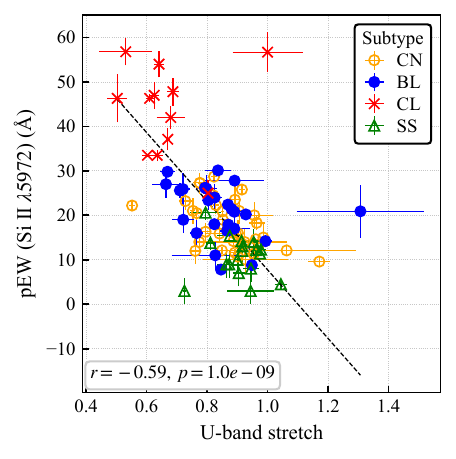}\\[-6pt]
    {\small (a) $U$-band stretch vs. pEW (Si II $\lambda5972$)}\\[4pt]
    \includegraphics[width=6.0cm]{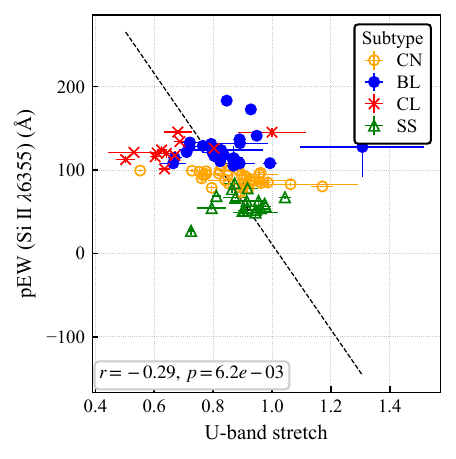}\\[-6pt]
    {\small (b) $U$-band stretch vs. pEW (Si II $\lambda6355$)}\\[4pt]
    \includegraphics[width=6.0cm]{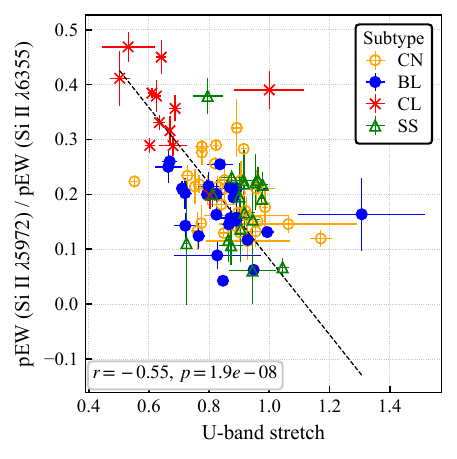}\\[-6pt]
    {\small (c) $U$-band stretch vs. pEW ratio of Si II $\lambda5972$ and $\lambda6355$}\\[4pt]
    \caption{Correlation between $U$-band stretch and the pEW of (a) Si II $\lambda5972$, (b) Si II $\lambda6355$, and (c) their ratio. The dashed line in each panel shows the best fit linear relation obtained with orthogonal distance regression. The Pearson coefficient $r$ and $p$-value are also shown. {Alt text: A figure consisting of three panels. All panels have U-band stretch as the x axis. The y axes show the pseudo-equivalent widths of the Si II line at 6355 Angstroms, the Si II line at 5972 Angstroms, and the ratio of the two lines, for each panel.}}
    \label{fig:su_pew}
\end{figure}

\begin{figure}[htbp]
    \centering
    \includegraphics[width=6.0cm]{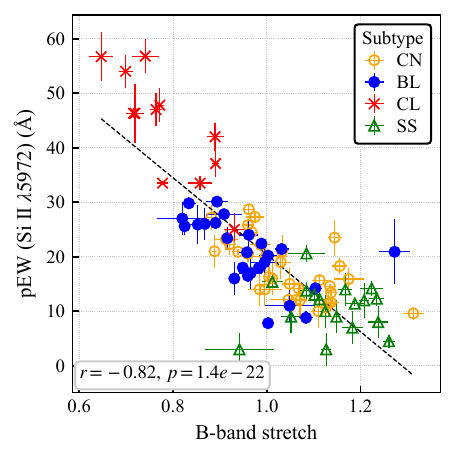}\\[-6pt]
    {\small (a) $B$-band stretch vs. pEW (Si II $\lambda5972$)}\\[4pt]
    \includegraphics[width=6.0cm]{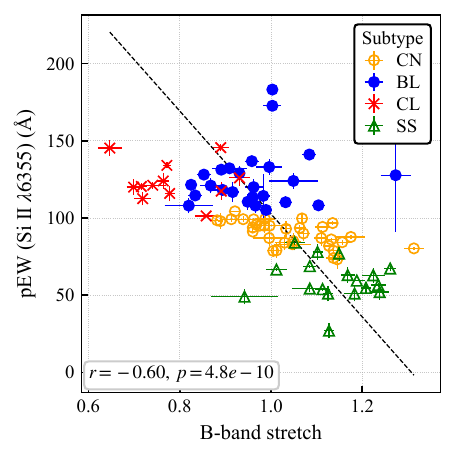}\\[-6pt]
    {\small (b) $B$-band stretch vs. pEW (Si II $\lambda6355$)}\\[4pt]
    \includegraphics[width=6.0cm]{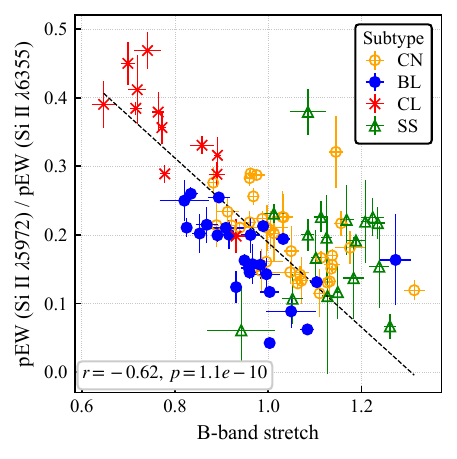}\\[-6pt]
    {\small (c) $B$-band stretch vs. pEW ratio of Si II $\lambda5972$ and $\lambda6355$}\\[4pt]
    \caption{The same as figure~\ref{fig:su_pew}, but for the $B$-band. {Alt text: A figure consisting of three panels. All panels have B-band stretch as the x axis. The y axes show the pseudo-equivalent widths of the Si II line at 6355 Angstroms, the Si II line at 5972 Angstroms, and the ratio of the two lines, for each panel.}}
    \label{fig:sb_pew}
\end{figure}

\begin{figure}[htbp]
    \centering
    \includegraphics[width=7.0cm]{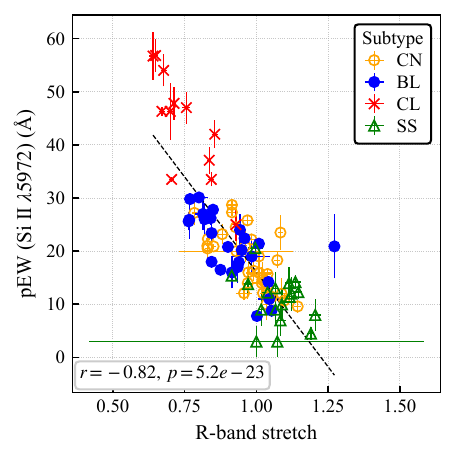}\\[-6pt]
    {\small (a) $R$-band stretch vs. pEW (Si II $\lambda5972$)}\\[4pt]
    \includegraphics[width=7.0cm]{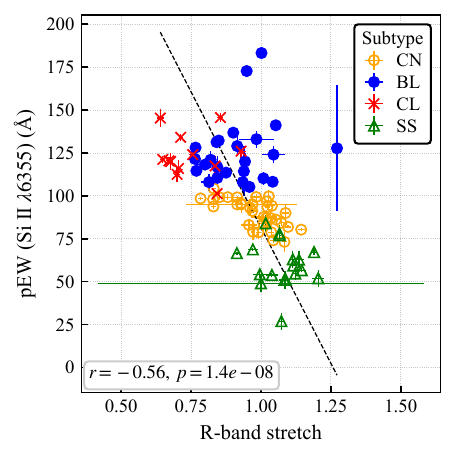}\\[-6pt]
    {\small (b) $R$-band stretch vs. pEW (Si II $\lambda6355$)}\\[4pt]
    \includegraphics[width=7.0cm]{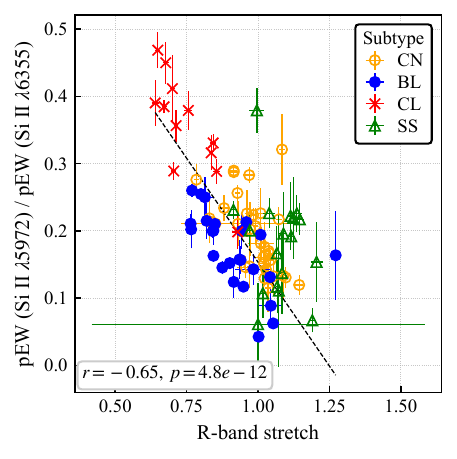}\\[-6pt]
    {\small (c) $R$-band stretch vs. pEW ratio of Si II $\lambda5972$ and $\lambda6355$}\\[4pt]
    \caption{The same as figure~\ref{fig:su_pew}, but for the $R$-band. {Alt text: A figure consisting of three panels. All panels have R-band stretch as the x axis. The y axes show the pseudo-equivalent widths of the Si II line at 6355 Angstroms, the Si II line at 5972 Angstroms, and the ratio of the two lines, for each panel.}}
    \label{fig:sr_pew}
\end{figure}

\begin{figure}[htbp]
    \centering
    \includegraphics[width=7.0cm]{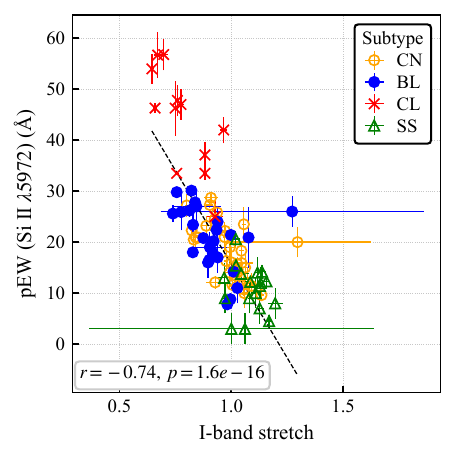}\\[-6pt]
    {\small (a) $I$-band stretch vs. pEW (Si II $\lambda5972$)}\\[4pt]
    \includegraphics[width=7.0cm]{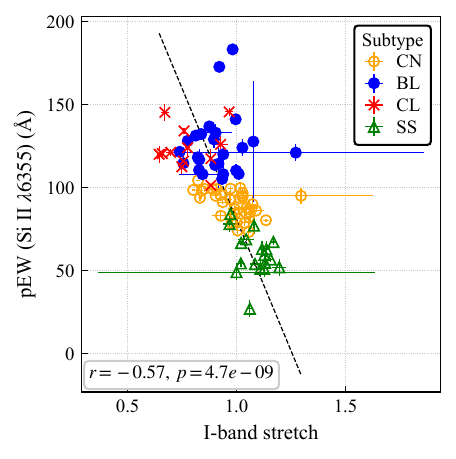}\\[-6pt]
    {\small (b) $I$-band stretch vs. pEW (Si II $\lambda6355$)}\\[4pt]
    \includegraphics[width=7.0cm]{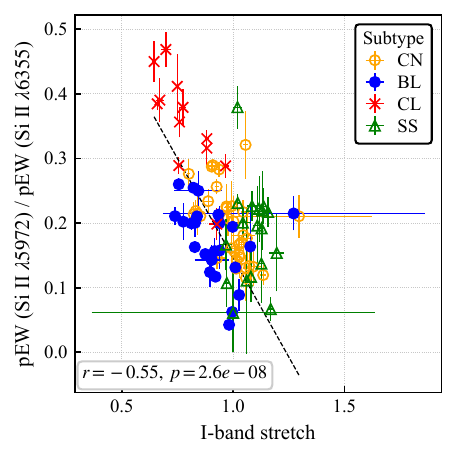}\\[-6pt]
    {\small (c) $I$-band stretch vs. pEW ratio of Si II $\lambda5972$ and $\lambda6355$}\\[4pt]
    \caption{The same as figure~\ref{fig:su_pew}, but for the $I$-band. {Alt text: A figure consisting of three panels. All panels have I-band stretch as the x axis. The y axes show the pseudo-equivalent widths of the Si II line at 6355 Angstroms, the Si II line at 5972 Angstroms, and the ratio of the two lines, for each panel.}}
    \label{fig:si_pew}
\end{figure}

This section compiles $U$, $B$, $R$, and $I$-band plots of stretch versus the pEWs of Si II $\lambda5972$, $\lambda6355$, and their ratio (figures~\ref{fig:su_pew}--\ref{fig:si_pew}). The $V$-band plots are presented in the main text (see subsection~\ref{ssec:fit_result}).

We also list the Pearson correlation coefficients ($r$) for these plots, when evaluated separately for each Branch subtype and band (table~\ref{tab:corr_pew}). 

\begin{table}[htbp]
  \tbl{Pearson $r$ and correlation significance based on $p$-values for stretch vs.\ three different pEW-related values by Branch subtype.}{%
  \begin{tabular}{lcccc}
      \hline
      Band & CN & BL & CL & SS \\
      \noalign{\vskip 2pt}
      \hline
      \noalign{\vskip 2pt}
      \multicolumn{5}{l}{\textbf{(1) pEW (Si II $\lambda$5972)}}\\
      $U$ & $-0.49^{*}$ & $-0.26^{\ddagger}$ & $-0.13^{\S}$ & $+0.22^{\S}$ \\
      $B$ & $-0.68^{*}$ & $-0.55^{*}$ & $-0.48^{*}$ & $+0.35^{\S}$ \\
      $V$ & $-0.66^{*}$ & $-0.67^{*}$ & $-0.52^{*}$ & $-0.21^{\S}$ \\
      $R$ & $-0.68^{*}$ & $-0.63^{*}$ & $-0.56^{*}$ & $-0.20^{\S}$ \\
      $I$ & $-0.56^{*}$ & $-0.47^{*}$ & $-0.58^{*}$ & $+0.07^{\S}$ \\
      \noalign{\vskip 2pt}
      \hline
      \noalign{\vskip 2pt}
      \multicolumn{5}{l}{\textbf{(2) pEW (Si II $\lambda$6355)}}\\
      $U$ & $-0.49^{\dagger}$  & $+0.02^{\S}$ & $-0.003^{\S}$ & $+0.47^{*}$ \\
      $B$ & $-0.63^{*}$ & $-0.40^{*}$ & $-0.03^{\S}$ & $+0.47^{*}$ \\
      $V$ & $-0.54^{*}$ & $-0.53^{*}$ & $-0.03^{\S}$ & $-0.13^{\S}$ \\
      $R$ & $-0.66^{*}$ & $-0.47^{*}$ & $-0.09^{\S}$ & $-0.03^{\S}$ \\
      $I$ & $-0.45^{*}$ & $-0.16^{\S}$ & $-0.14^{\S}$ & $+0.08^{\S}$ \\
      \noalign{\vskip 2pt}
      \hline
      \noalign{\vskip 2pt}
      \multicolumn{5}{l}{\textbf{(3) pEW (Si II $\lambda$5972) / pEW (Si II $\lambda$6355)}}\\
      $U$ & $-0.45^{*}$ & $-0.35^{\dagger}$ & $-0.33^{\dagger}$ & $-0.22^{\ddagger}$ \\
      $B$ & $-0.53^{*}$ & $-0.59^{*}$ & $-0.80^{*}$ & $+0.04^{\S}$ \\
      $V$ & $-0.52^{*}$ & $-0.58^{*}$ & $-0.83^{*}$ & $-0.16^{\S}$ \\
      $R$ & $-0.52^{*}$ & $-0.60^{*}$ & $-0.82^{*}$ & $-0.22^{\S}$ \\
      $I$ & $-0.46^{*}$ & $-0.37^{*}$ & $-0.79^{*}$ & $-0.003^{\S}$ \\
      \noalign{\vskip 2pt}
      \hline
    \end{tabular}}\label{tab:corr_pew}
\begin{tabnote}
\par\noindent
\footnotemark[$*$] $p<{10}^{-3}$.
\par\noindent
\footnotemark[$\dagger$] ${10}^{-3}\leq p<{10}^{-2}$. 
\par\noindent
\footnotemark[$\ddagger$] ${10}^{-2}\leq p<0.05$.
\par\noindent
\footnotemark[$\S$] $0.05\leq p$.
\end{tabnote}
\end{table}

\section{Individual light curves used to construct the average templates}\label{app:individual_lc}
\begin{figure*}[t]
    \centering
    % --- U-band (row 1) ---
    \begin{subfigure}{0.24\textwidth}
        \includegraphics[width=\linewidth]{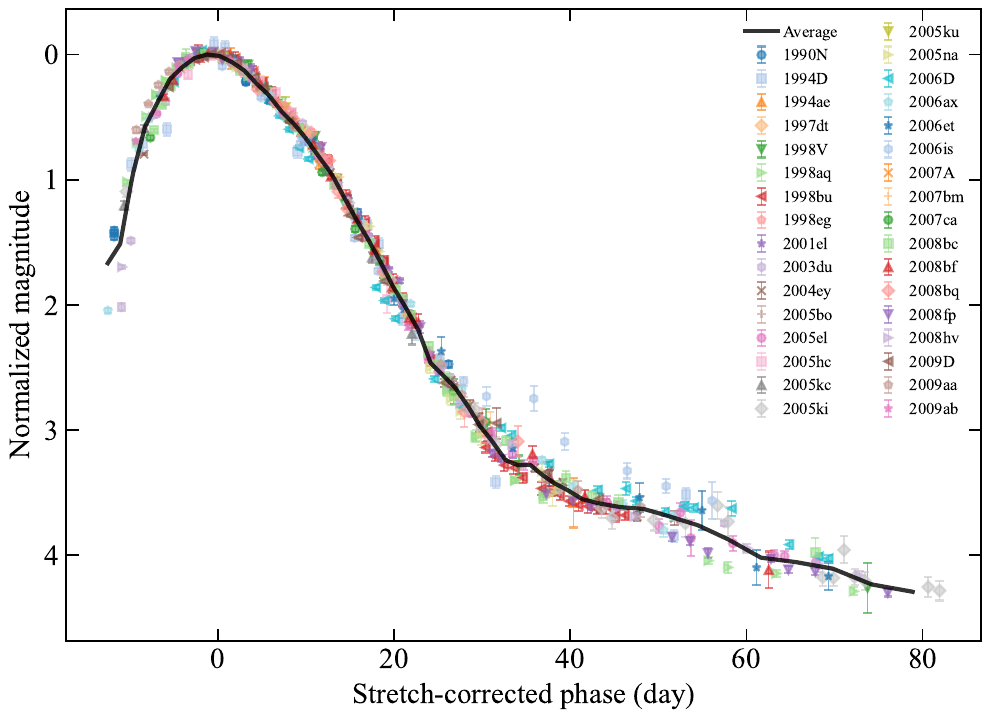}
        \caption*{CN ($U$-band)}
    \end{subfigure}
    \begin{subfigure}{0.24\textwidth}
        \includegraphics[width=\linewidth]{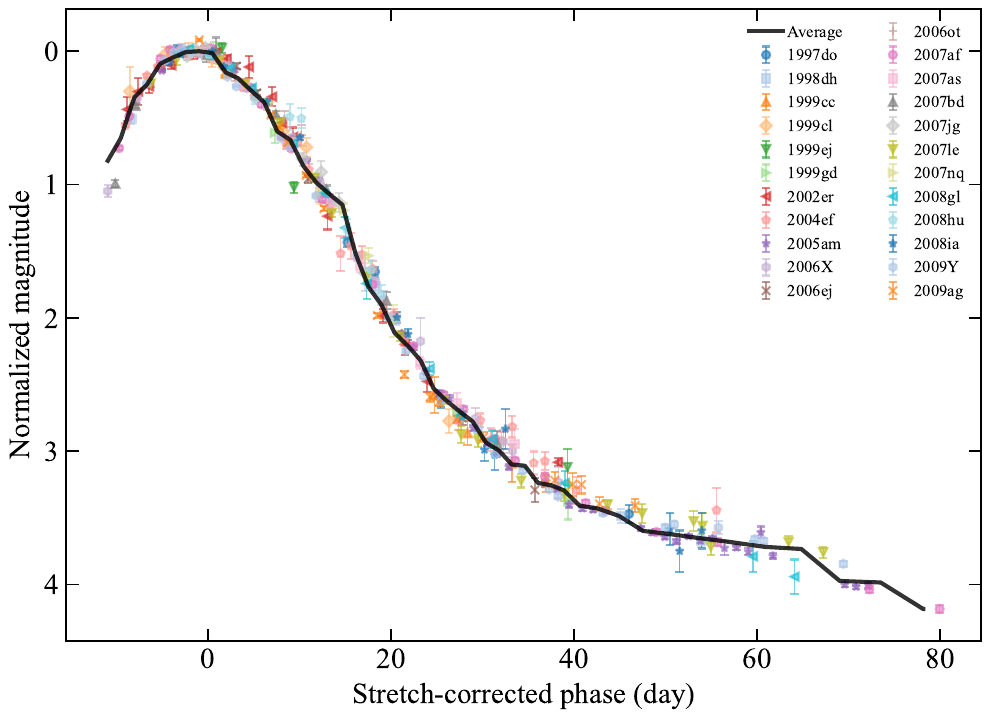}
        \caption*{BL ($U$-band)}
    \end{subfigure}
    \begin{subfigure}{0.24\textwidth}
        \includegraphics[width=\linewidth]{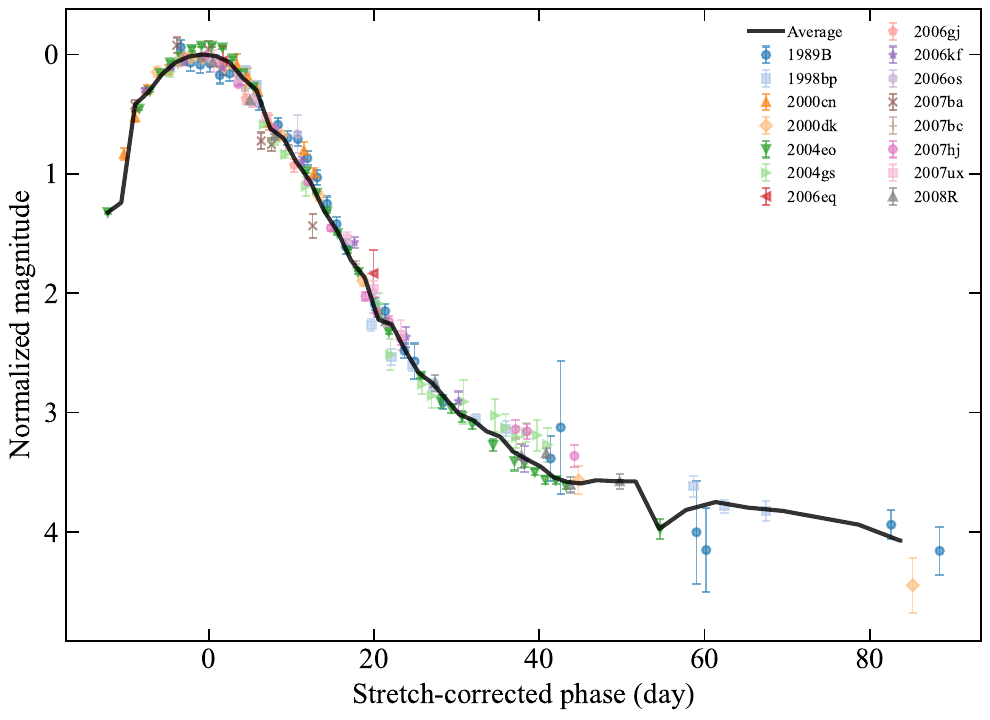}
        \caption*{CL ($U$-band)}
    \end{subfigure}
    \begin{subfigure}{0.24\textwidth}
        \includegraphics[width=\linewidth]{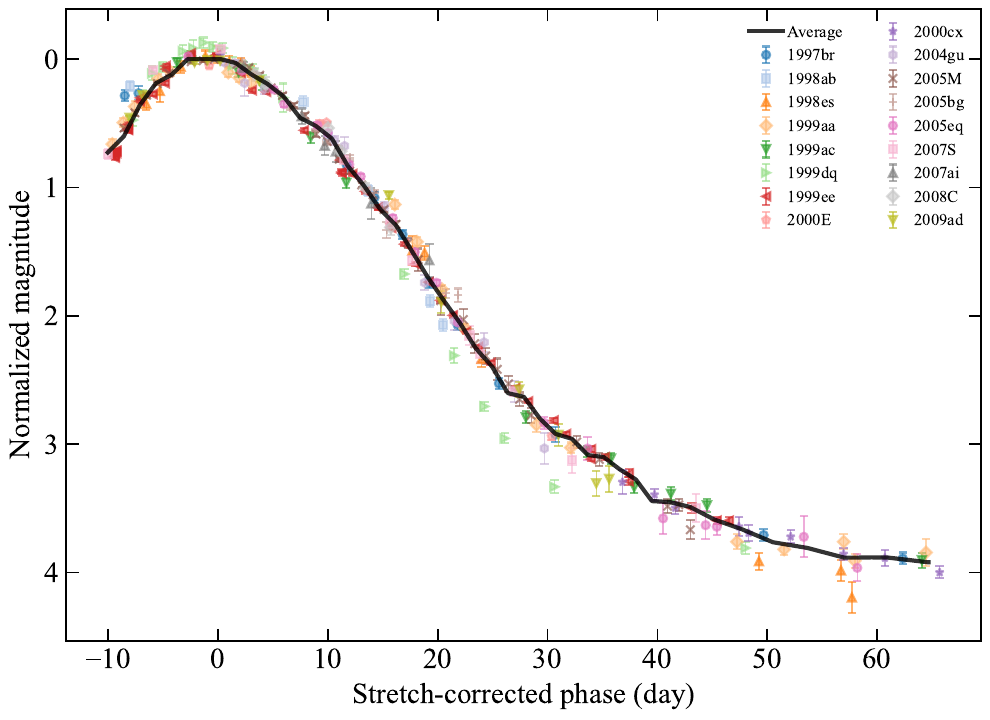}
        \caption*{SS ($U$-band)}
    \end{subfigure}

    % --- B-band (row 2) ---
    \begin{subfigure}{0.24\textwidth}
        \includegraphics[width=\linewidth]{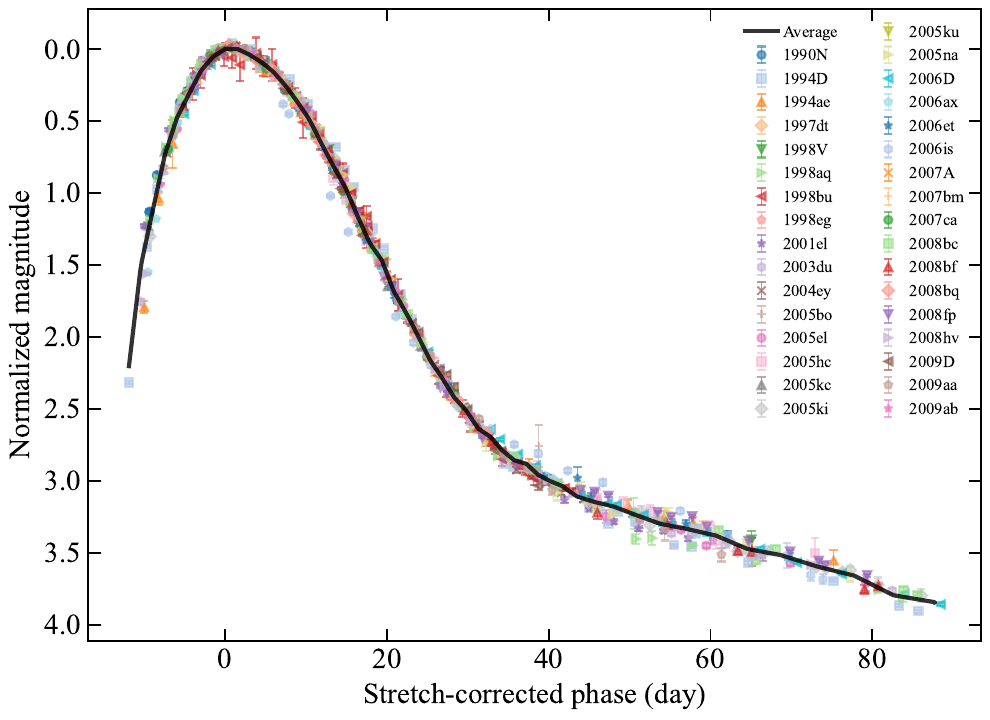}
        \caption*{CN ($B$-band)}
    \end{subfigure}
    \begin{subfigure}{0.24\textwidth}
        \includegraphics[width=\linewidth]{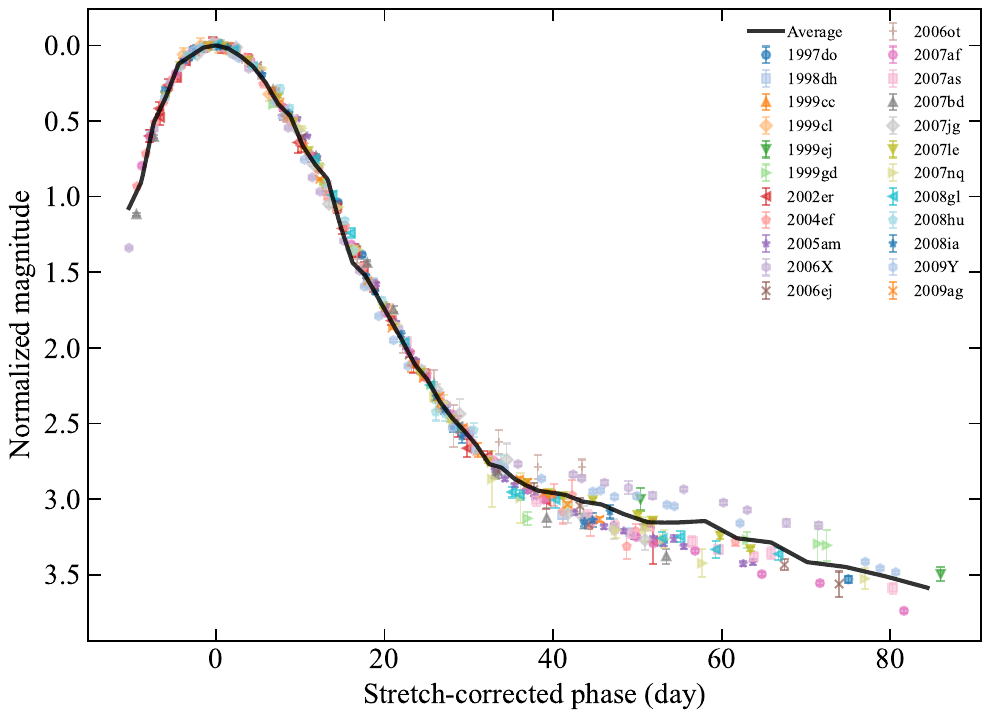}
        \caption*{BL ($B$-band)}
    \end{subfigure}
    \begin{subfigure}{0.24\textwidth}
        \includegraphics[width=\linewidth]{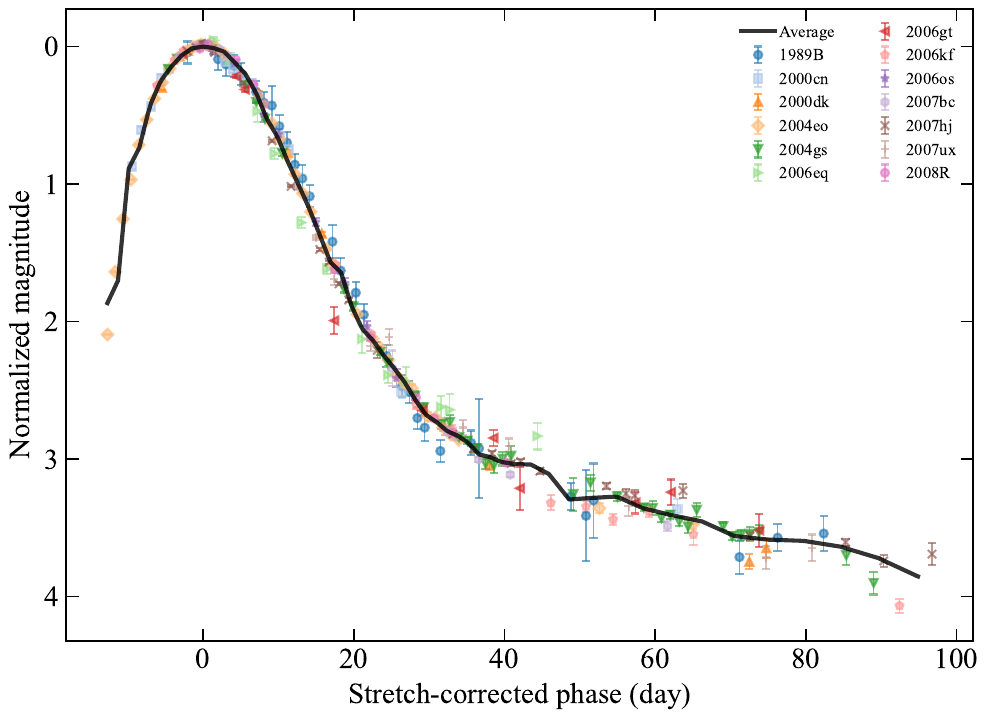}
        \caption*{CL ($B$-band)}
    \end{subfigure}
    \begin{subfigure}{0.24\textwidth}
        \includegraphics[width=\linewidth]{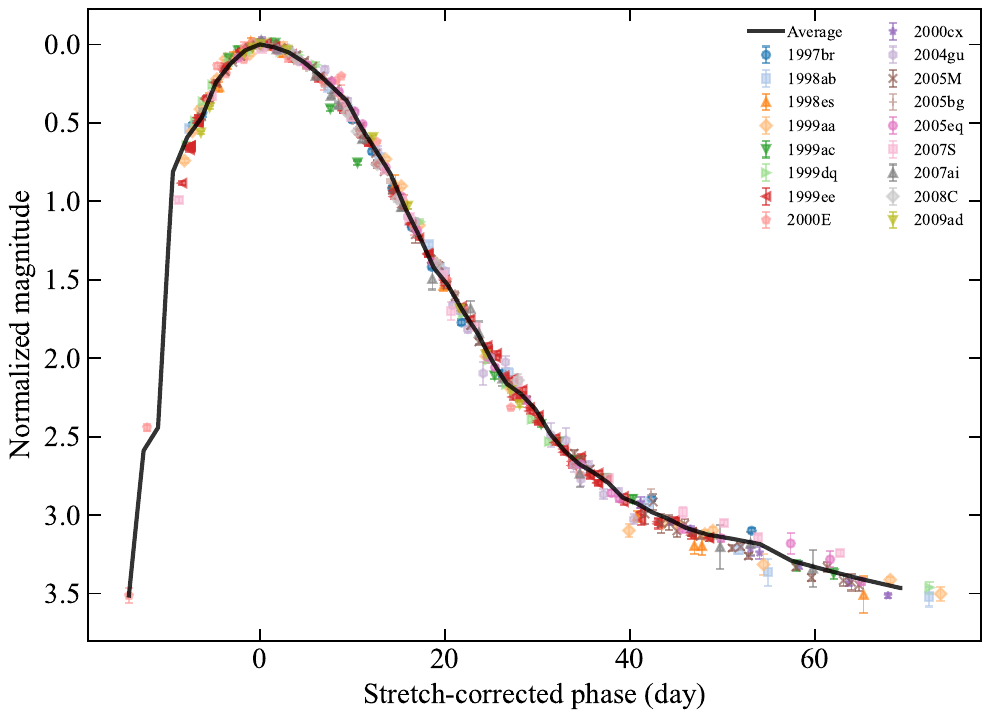}
        \caption*{SS ($B$-band)}
    \end{subfigure}

    % --- V-band (row 3) ---
    \begin{subfigure}{0.24\textwidth}
        \includegraphics[width=\linewidth]{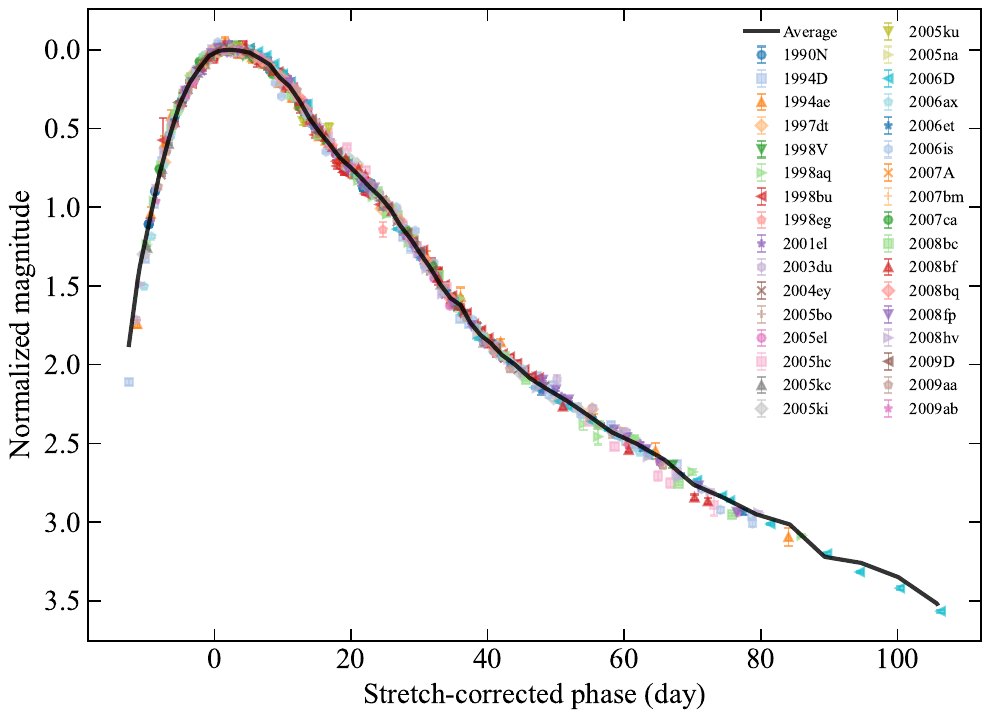}
        \caption*{CN ($V$-band)}
    \end{subfigure}
    \begin{subfigure}{0.24\textwidth}
        \includegraphics[width=\linewidth]{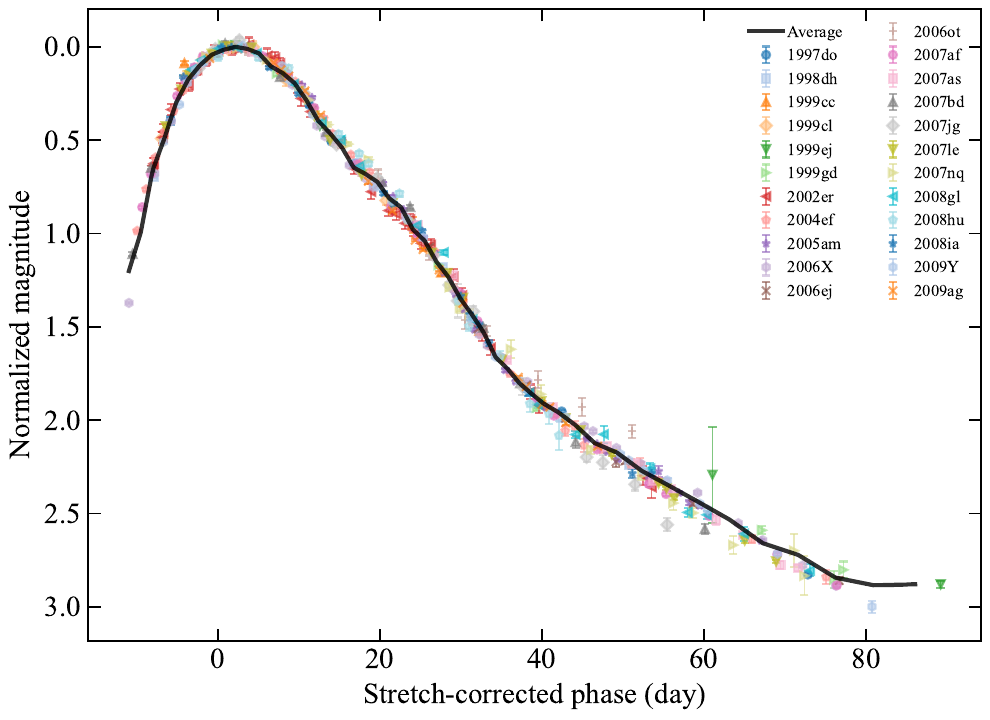}
        \caption*{BL ($V$-band)}
    \end{subfigure}
    \begin{subfigure}{0.24\textwidth}
        \includegraphics[width=\linewidth]{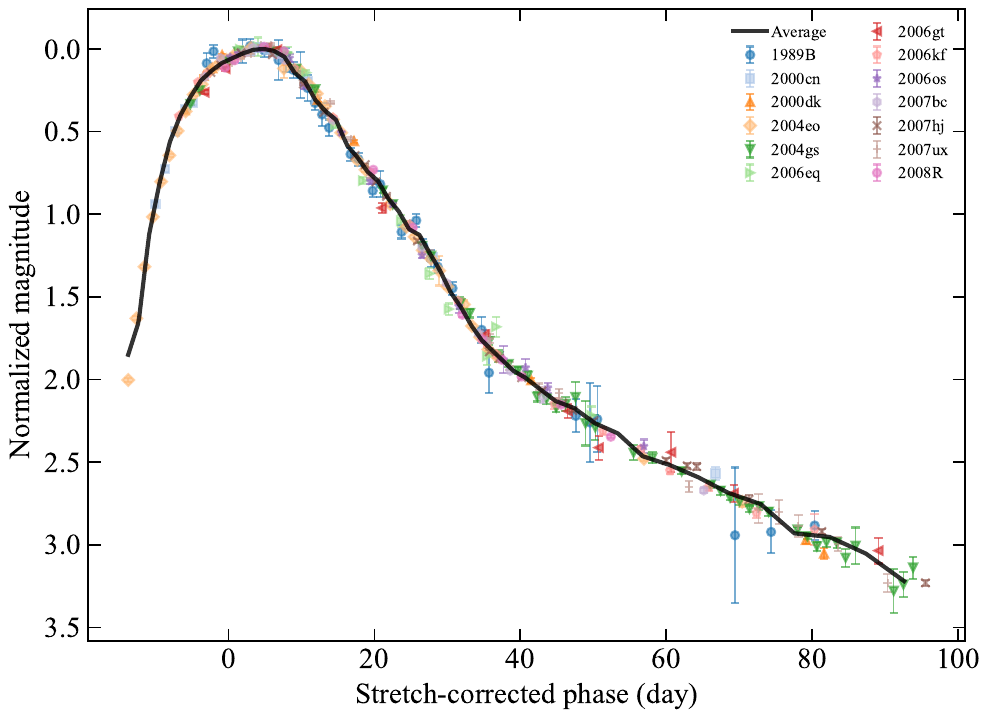}
        \caption*{CL ($V$-band)}
    \end{subfigure}
    \begin{subfigure}{0.24\textwidth}
        \includegraphics[width=\linewidth]{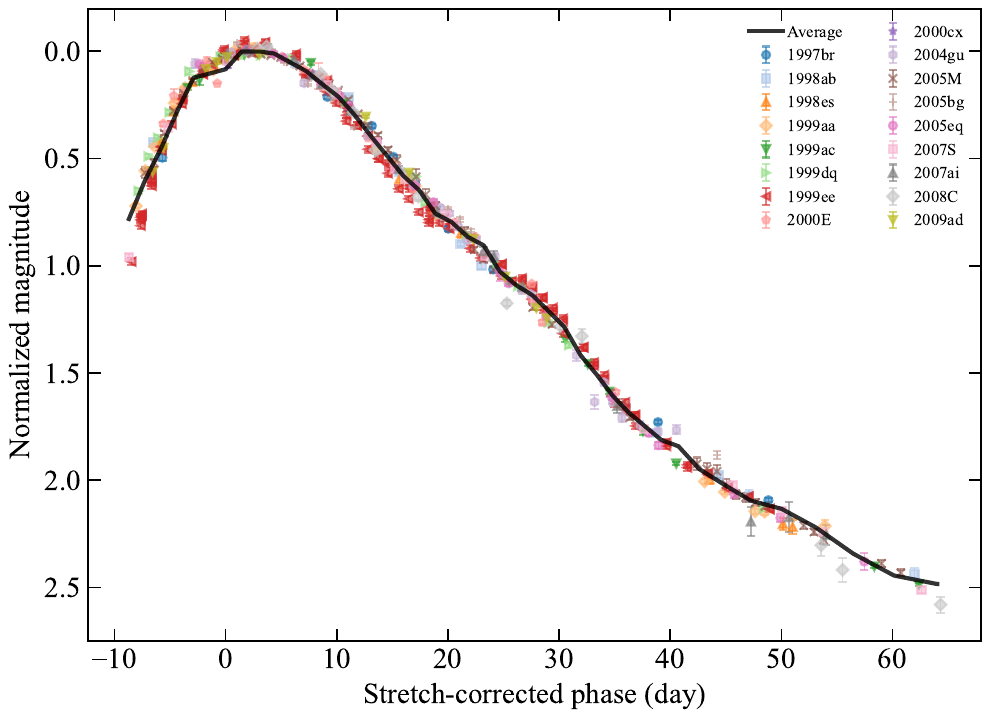}
        \caption*{SS ($V$-band)}
    \end{subfigure}

    % --- R-band (row 4) ---
    \begin{subfigure}{0.24\textwidth}
        \includegraphics[width=\linewidth]{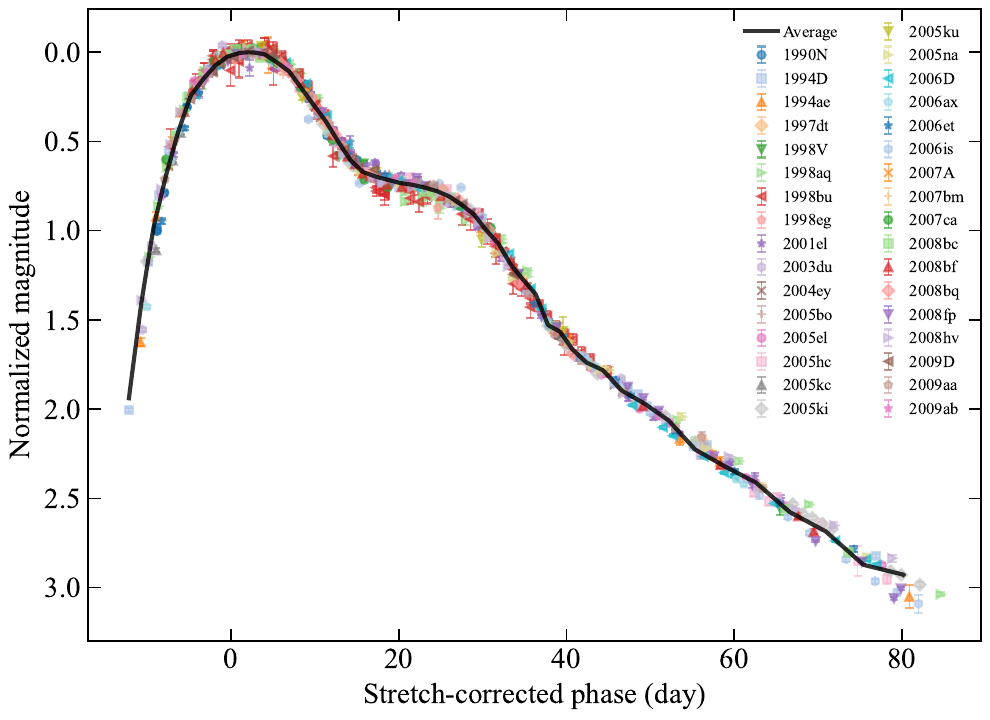}
        \caption*{CN ($R$-band)}
    \end{subfigure}
    \begin{subfigure}{0.24\textwidth}
        \includegraphics[width=\linewidth]{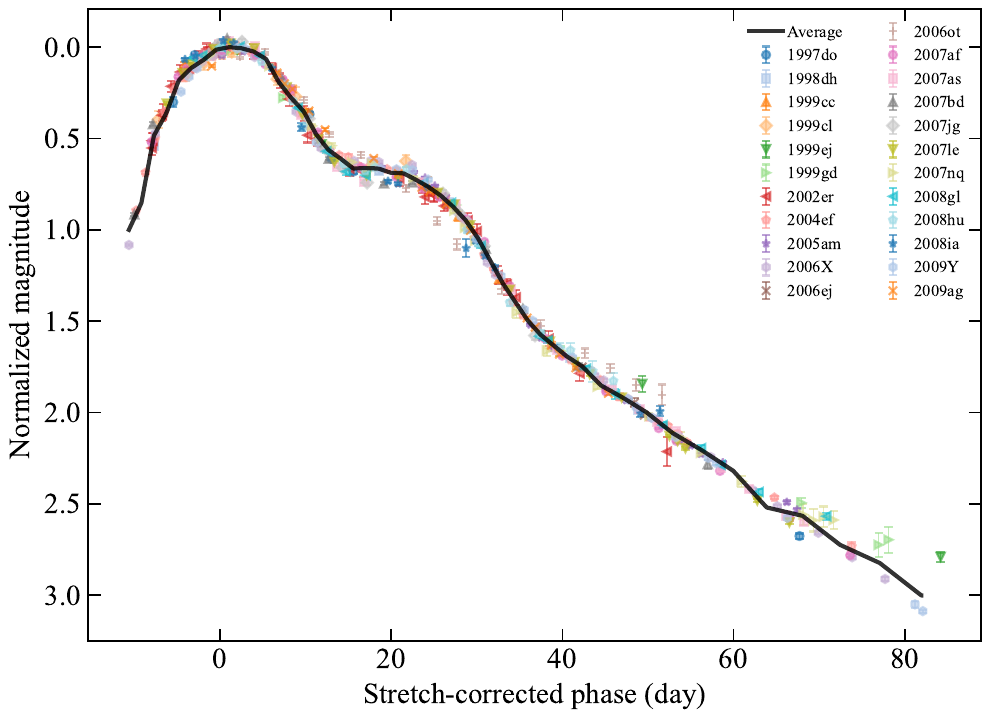}
        \caption*{BL ($R$-band)}
    \end{subfigure}
    \begin{subfigure}{0.24\textwidth}
        \includegraphics[width=\linewidth]{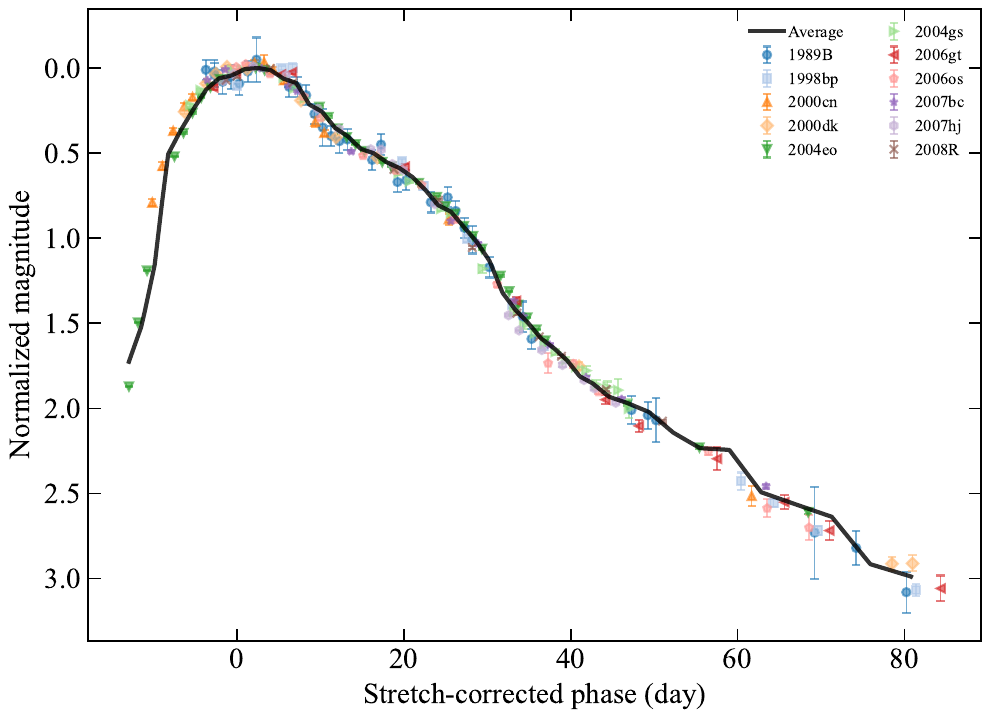}
        \caption*{CL ($R$-band)}
    \end{subfigure}
    \begin{subfigure}{0.24\textwidth}
        \includegraphics[width=\linewidth]{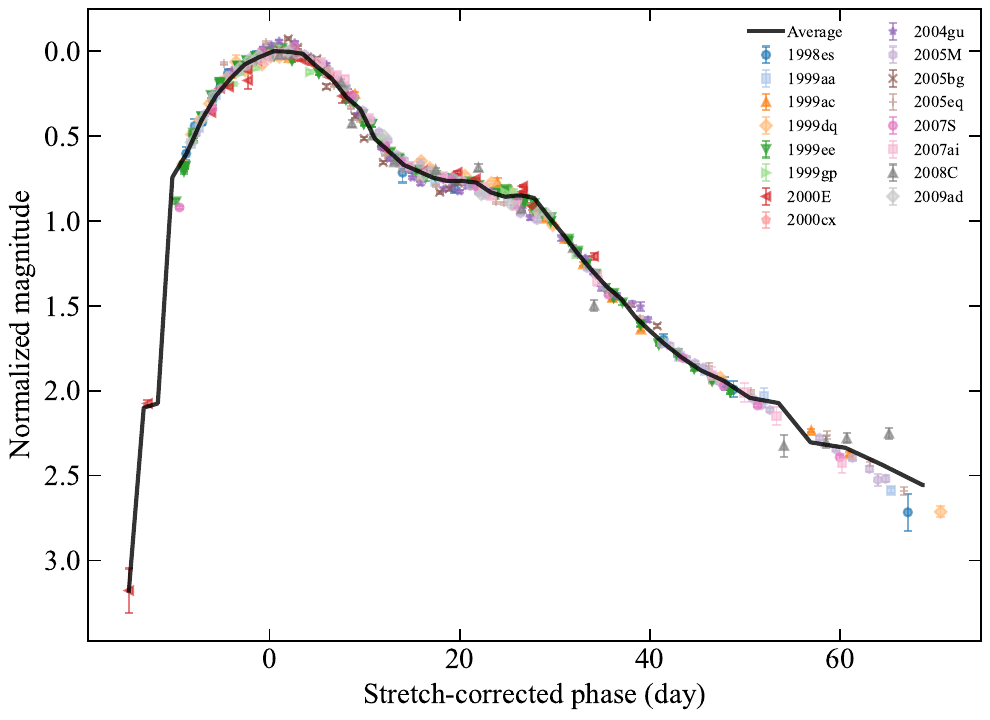}
        \caption*{SS ($R$-band)}
    \end{subfigure}

    % --- I-band (row 5) ---
    \begin{subfigure}{0.24\textwidth}
        \includegraphics[width=\linewidth]{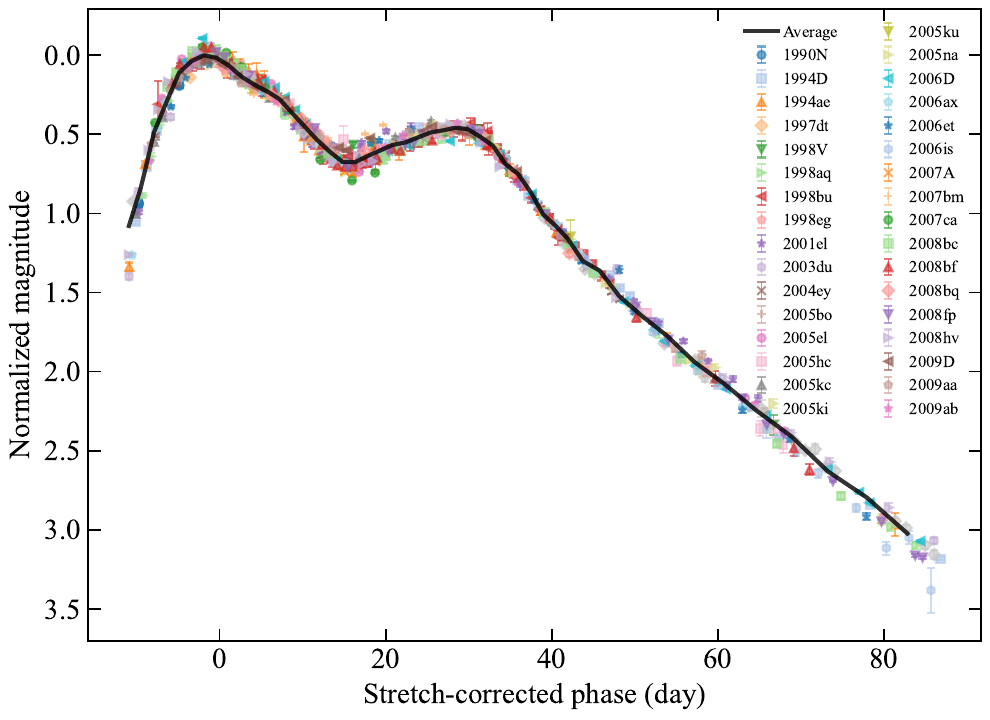}
        \caption*{CN ($I$-band)}
    \end{subfigure}
    \begin{subfigure}{0.24\textwidth}
        \includegraphics[width=\linewidth]{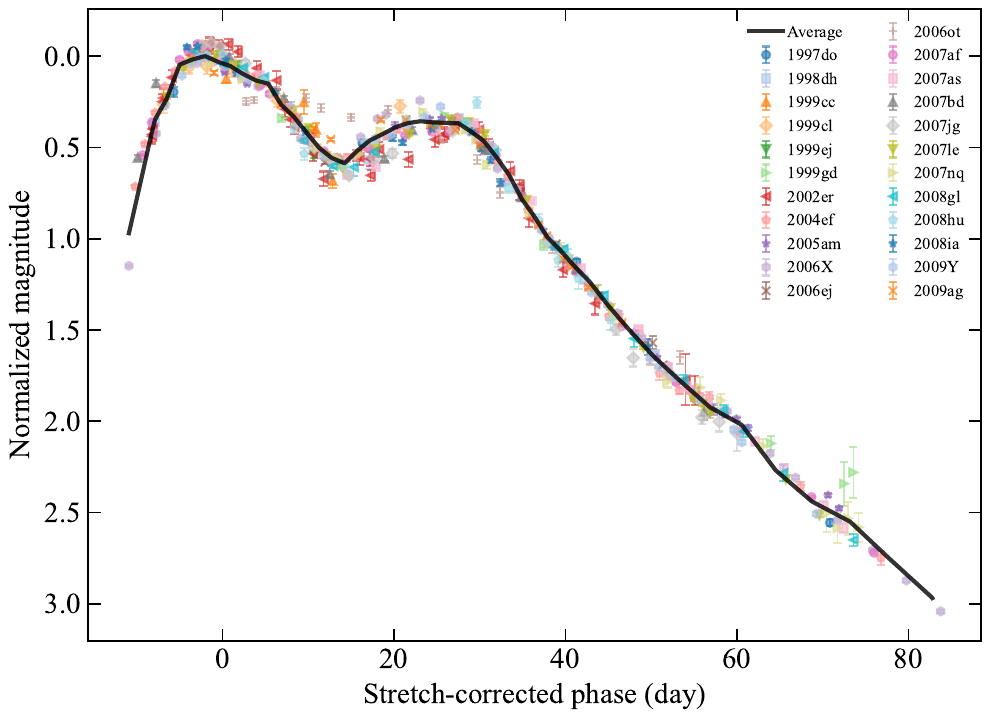}
        \caption*{BL ($I$-band)}
    \end{subfigure}
    \begin{subfigure}{0.24\textwidth}
        \includegraphics[width=\linewidth]{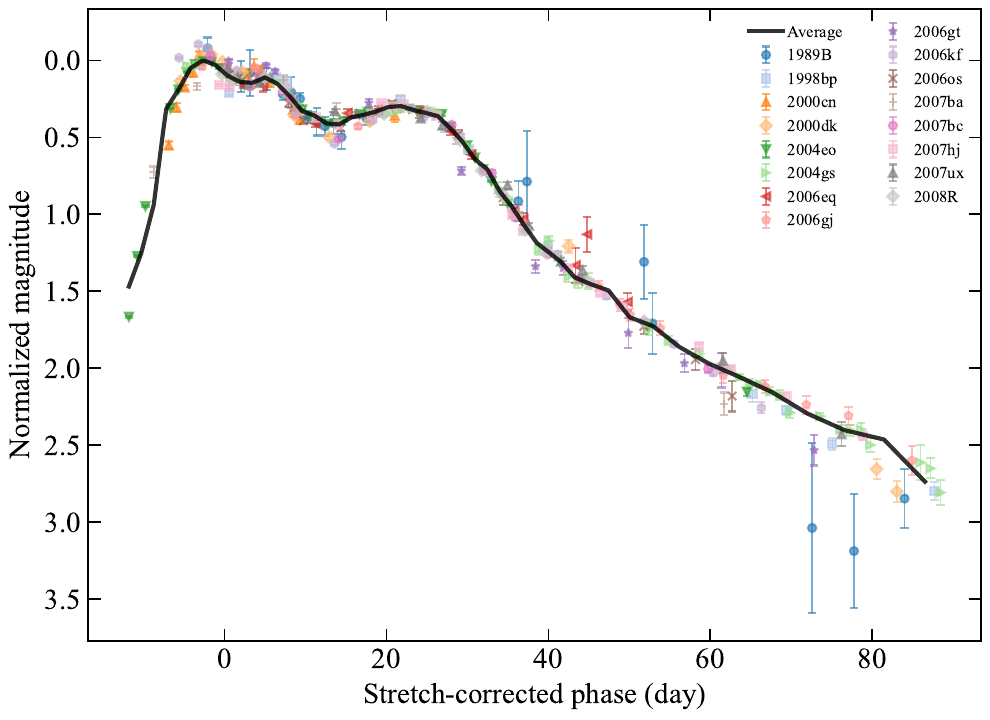}
        \caption*{CL ($I$-band)}
    \end{subfigure}
    \begin{subfigure}{0.24\textwidth}
        \includegraphics[width=\linewidth]{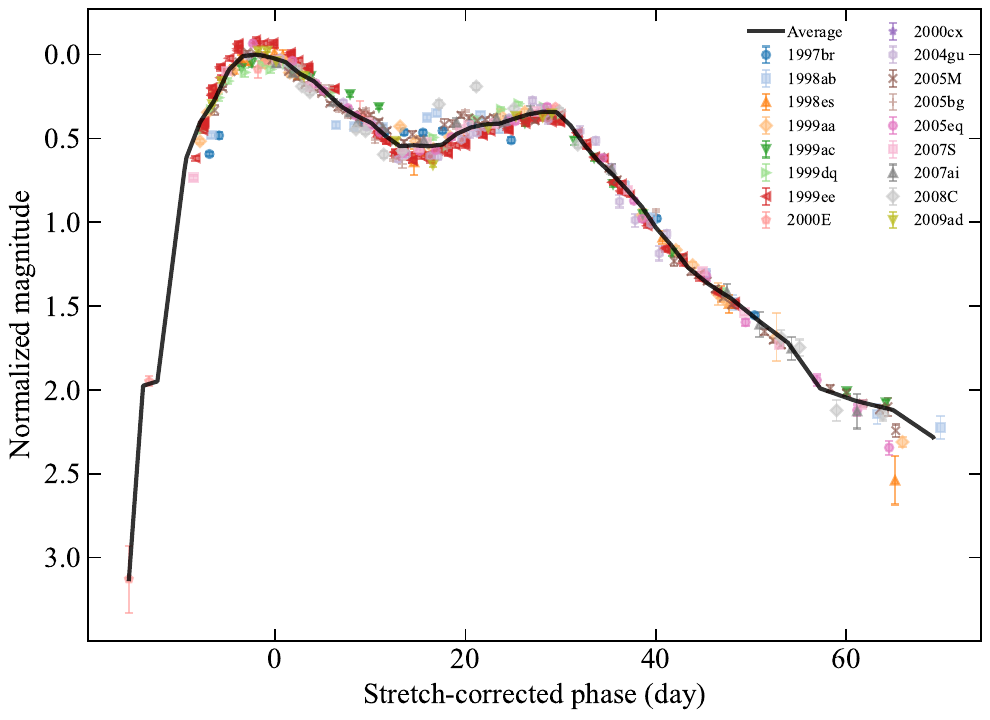}
        \caption*{SS ($I$-band)}
    \end{subfigure}

    \caption{Individual $UBVRI$ light curves used to construct the averaged templates for each Branch subtype. Each panel shows all fitted light curves with the resulting averaged template overplotted (black solid line). Rows correspond to photometric bands ($U$, $B$, $V$, $R$, $I$), and columns correspond to Branch subtypes (CN, BL, CL, SS). These plots illustrate the data coverage and intrinsic dispersion within each subgroup. {Alt text: A 5×4 grid showing all Branch subtypes (CN, BL, CL, SS) in columns and photometric bands (U, B, V, R, I) in rows. Colored markers show the individual light curves, and solid black lines show the averaged light curves.}}
    \label{fig:individual_lc}
\end{figure*}

This section presents the individual $UBVRI$ light curves after the first iteration (fitting to the TAK08 template) used to construct the average light curves shown in figures~\ref{fig:average_lc}. In figure~\ref{fig:individual_lc}, all fitted light curves are overplotted in a single panel for each band and subtype.

% Any journal's BST file (e.g., apj.bst) can be used as PASJ's BST is unavailable.    
% \bibliographystyle{****}
% \bibliography{****}

\end{document}